\documentclass[acmsmall]{acmart}

\AtBeginDocument{%
  \providecommand\BibTeX{{%
    \normalfont B\kern-0.5em{\scshape i\kern-0.25em b}\kern-0.8em\TeX}}}

\setcopyright{acmcopyright}
\copyrightyear{2018}
\acmYear{2018}
\acmDOI{XXXXXXX.XXXXXXX}

\acmJournal{JACM}
\acmVolume{37}
\acmNumber{4}
\acmArticle{111}
\acmMonth{8}

\usepackage{multirow}
\usepackage[figuresright]{rotating}
\usepackage{graphicx}
\usepackage{colortbl}
\usepackage{pifont}

\makeatletter
\usepackage{comment}
\let\wfs@comment@comment\comment
\let\comment\@undefined
\usepackage{changes}
\let\wfs@changes@comment\comment
\let\comment\@undefined
\newcommand\comment{%
    \ifthenelse{\equal{\@currenvir}{comment}}
    {\wfs@comment@comment}
    {\wfs@changes@comment}%
}
\makeatother

\begin{document}

\title{Machine Unlearning: A Survey}


\author{Heng Xu}
\email{Heng.Xu-2@student.uts.edu.au}

\author{Tianqing Zhu*}
\email{Tianqing.Zhu@uts.edu.au}

\author{Lefeng Zhang}
\email{Lefeng.Zhang@student.uts.edu.au}
\affiliation{
  \institution{University of Technology Sydney}
  \streetaddress{123 Broadway}
  \city{Ultimo NSW 2007}
  \country{Australia}
}

\author{Wanlei Zhou}
\email{wlzhou@cityu.edu.mo}
\affiliation{%
  \institution{City University of Macau}
  \streetaddress{Avenida Padre Tomás Pereira Taipa}
  \state{Macau 999078}
  \country{China}
}

\author{Philip S. Yu}
\email{psyu@cs.uic.edu}
\affiliation{%
  \institution{University of Illinois at Chicago}
  \streetaddress{1200 W Harrison St, Chicago}
  \state{Illinois 60607}
  \country{United States}
}

\renewcommand{\shortauthors}{H. Xu et al.}

\begin{abstract}
	Machine learning has attracted widespread attention and evolved into an enabling technology for a wide range of highly successful applications, such as intelligent computer vision, speech recognition, medical diagnosis, and more. Yet a special need has arisen where, due to privacy, usability, and/or \textit{the right to be forgotten}, information about some specific samples needs to be removed from a model, called machine unlearning. This emerging technology has drawn significant interest from both academics and industry due to its innovation and practicality. At the same time, this ambitious problem has led to numerous research efforts aimed at confronting its challenges. To the best of our knowledge, no study has analyzed this complex topic or compared the feasibility of existing unlearning solutions in different kinds of scenarios. Accordingly, with this survey, we aim to capture the key concepts of unlearning techniques. The existing solutions are classified and summarized based on their characteristics within an up-to-date and comprehensive review of each category's advantages and limitations. The survey concludes by highlighting some of the outstanding issues with unlearning techniques, along with some feasible directions for new research opportunities. 
	
\end{abstract}



\maketitle

\section{Introduction}

In recent years, machine learning has seen remarkable progress and wide exploration across every field of artificial intelligence (AI)~\cite{doi:10.1126/science.aaa8415}. However, as AI becomes increasingly data-dependent, more and more factors, such as privacy concerns, regulations and laws, are leading to a new type of request – to delete information. Specifically, concerned parties are requesting that particular samples be removed from a training dataset and that the impact of those samples be removed from an already-trained model~\cite{DBLP:conf/uss/Carlini0EKS19,DBLP:conf/ccs/ZhangRWRCHZ21,DBLP:journals/tifs/KhosravyNHNB22}. This is because membership inference attacks~\cite{DBLP:conf/sp/ShokriSSS17} and model inversion attacks~\cite{DBLP:conf/ccs/FredriksonJR15} can reveal information about the specific contents of a training dataset. More importantly, legislators around the world have wisely introduced laws that grant users \textit{the right to be forgotten}~\cite{DBLP:journals/clsr/VillarongaKL18,DBLP:journals/corr/abs-1807-04644}. These regulations, which include the European Union’s General Data Protection Regulation (GDPR)~\cite{webpage:GDPR}, the California Consumer Privacy Act (CCPA)~\cite{webpage:CCPA}, the Act on the Protection of Personal Information (APPI)~\cite{webpage:APPI}, and Canada’s proposed Consumer Privacy Protection Act (CPPA)~\cite{webpage:CPPA}, compel the deletion of private information.

\subsection{The Motivation of machine unlearning}

\begin{figure}
  \centering
  \includegraphics[width=0.6\linewidth]{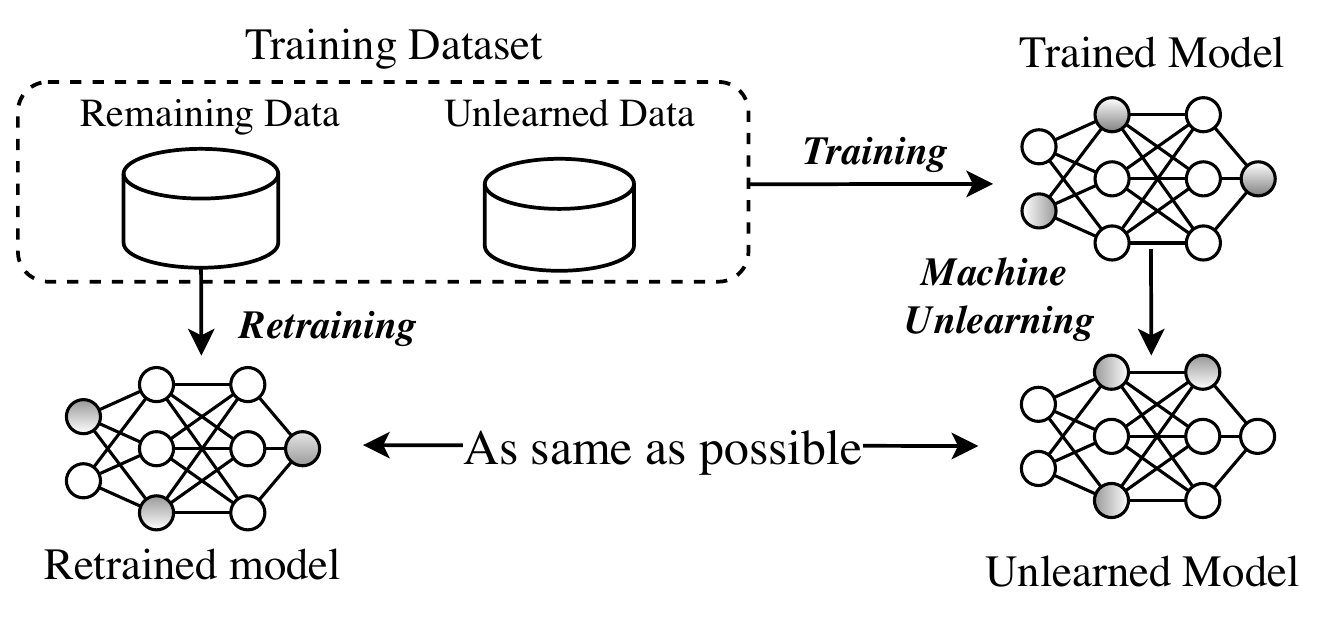}
  \caption{Illustration of Machine Unlearning.}
  \label{figure:machinunlearning}
\end{figure}

Machine unlearning (a.k.a. selectively forgetting, data deletion, or scrubbing) requires that the samples and their influence can be completely and quickly removed from a training dataset and a trained model~\cite{DBLP:conf/nips/GinartGVZ19,DBLP:conf/nips/NguyenLJ20,DBLP:conf/eurocrypt/GargGV20}. Figure~\ref{figure:machinunlearning} illustrates an example of machine unlearning for a trained model. 

Machine unlearning is not only motivated by regulations and laws. It also stems from the privacy and security concerns of the data provider, as well as the requirement of model owners themselves. In fact, removing the influence of outlier training samples from a model will lead to higher model performance and robustness~\cite{DBLP:conf/sigmod/Krishnan017}. There are existing data protection techniques that are similar to machine unlearning, but they differ in either objectives or rationales.

Here, we briefly discuss the
main differences between current techniques and machine unlearning.

\begin{itemize}
    \item \textbf{Differential Privacy.} Differential privacy~\cite{DBLP:journals/tkde/zhu2020,DBLP:journals/jsac/WeiLMDCJHP22} guarantees that by looking at a model output, one cannot tell whether a sample is in the training dataset or not. This technique ensures a subtle bound on the contribution of \emph{every} sample to the final model~\cite{DBLP:journals/tkde/ZhuLZY17,DBLP:journals/csur/ZhangZXZY22}, but machine unlearning is targeted on the removing of \emph{user-specific} training samples.

    \item \textbf{Data Masking.} Data masking~\cite{DBLP:journals/eswa/XiangZDXYT21} is designed to hide sensitive information in the original dataset. It transforms sensitive data to prevent them from being disclosed in unreliable environments~\cite{DBLP:conf/trustcom/WangWJTZLL21}. 
    In comparison, The objective of machine unlearning is to prevent a trained model from leaking sensitive information about its training samples. 

    \item \textbf{Online Learning.} Online learning~\cite{DBLP:journals/ai/Al-OmariLHC22} adjusts models quickly according to the data in a feedback process, such that the model can reflect online changes in a timely manner. One major difference between online learning and machine unlearning is that the former requires a merge operation to incorporate updates, while machine unlearning is an inverse operation that eliminates those updates when an unlearning request is received~\cite{DBLP:conf/cidr/Schelter20}. 

    \item \textbf{Catastrophic forgetting.} Catastrophic forgetting~\cite{DBLP:conf/icml/Chen0HD21,DBLP:conf/aaai/LiuYW21} refers to a significant drop in performance on previously learned tasks when a model is fine-tuned for a new task. Catastrophic forgetting causes a deep network to lose accuracy, but the information of the data it uses may still be accessible by analyzing the weights~\cite{DBLP:journals/corr/abs-2111-08947}, Therefore, it does not satisfy the conditions required by machine unlearning. 

\end{itemize}

When users revoke permissions over some training data, it is not sufficient to merely remove those data from the original training dataset, since the attackers can still reveal user information from the trained models~\cite{DBLP:conf/colt/UllahM0RA21}. One straightforward approach to perfectly removing  information from the model is to retrain it from scratch (the retraining process in Figure~\ref{figure:machinunlearning}). However, many complex models have been built on an enormous set of samples. Retraining is generally a computationally expensive process~\cite{DBLP:conf/sp/CaoY15,DBLP:conf/sp/BourtouleCCJTZL21}. Moreover, in some specific learning scenarios, such as federated learning~\cite{DBLP:journals/corr/abs-2201-09441,DBLP:journals/tist/YangLCT19}, the training dataset may not be accessible, and thus retraining cannot be conducted at all. Therefore, to reduce the computational cost and make machine unlearning possible in all circumstances, 
new techniques should be proposed (the unlearning process in Figure~\ref{figure:machinunlearning}). 


\subsection{Contributions of this survey}


Machine unlearning has played an essential role in many applications~\cite{DBLP:conf/www/Chen0ZD22,DBLP:journals/corr/abs-2203-11491}. However, its implementation and verification strategies are still not fully explored. There are various concepts and multiple verification schemes in this field, and the boundary between machine unlearning and other techniques is vague. These phenomena motivate us to compile a comprehensive survey that summarizes, analyzes, and categorizes machine unlearning techniques. In this survey, we aim to find a clear way to present the ideas and concepts in machine unlearning, showing their characteristics and highlighting their advantage. In addition, we propose a novel taxonomy for classifying state-of-the-art literature. We hope this survey provides an in-depth overview to readers who wish to know this field, and it also serves as a stepping-stone for advancing innovations and widening research visions. The main contributions of this paper are listed as follows:

\begin{itemize}
\item We proposed a novel taxonomy of current machine unlearning techniques based on their rationale and unlearning strategy.
\item We comprehensively summarized state-of-the-art unlearning methods based on the proposed taxonomy, showing their benefits and shortcomings. 
\item We summarized the verification methods of machine unlearning within the taxonomy, and reviewed their implementations with related unlearning techniques. 
	\item We provided critical and deep discussions on the open issues in machine unlearning, as well as pointing out possible further research directions.
\end{itemize}

\subsection{Comparison to existing surveys in machine unlearning} 

There are some works that have been conducted to summarize machine unlearning. However, few of them provide deep and comprehensive insight into current research. Here we introduce some relevant works for reference. Table~\ref{tab:comparison} summarizes the comparison of those references.

\begin{table}[]
\centering
\caption{Comparison between Existing Machine Unlearning Surveys.}
\renewcommand{\arraystretch}{1.2}
\label{tab:comparison}
\resizebox{\textwidth}{!}{
\begin{tabular}{c|cccc|ccc|cccccc|cccccc|cccc}
\hline
\rowcolor{gray!40}
 Survey & \multicolumn{4}{c|}{Targets} & \multicolumn{3}{c|}{Desiderata} & \multicolumn{6}{c|}{Unlearning Request}   & \multicolumn{6}{c|}{Verification Methods}   & \multicolumn{4}{c}{Open Questions}        \\ \cline{2-24} 
    
& \rotatebox{90}{Exact}  & \rotatebox{90}{Approximate}  & \rotatebox{90}{Strong}  & \rotatebox{90}{Weak} & \rotatebox{90}{Consistency}  & \rotatebox{90}{Accuracy}  & \rotatebox{90}{Verifiability}  & \rotatebox{90}{Sample} & \rotatebox{90}{Class} & \rotatebox{90}{Feature} & \rotatebox{90}{Sequence} & \rotatebox{90}{Graph} & \rotatebox{90}{Client} & \rotatebox{90}{Retraining-based}  & \rotatebox{90}{Attack-based} & \rotatebox{90}{Accuracy-based} & \rotatebox{90}{Relearn Time-based} & \rotatebox{90}{Theory-based} & \rotatebox{90}{Information bound-based} & \rotatebox{90}{Universality} & \rotatebox{90}{Security} & \rotatebox{90}{Verification} & \rotatebox{90}{Applications} \\ \hline

\rowcolor{gray!40}
Thanh et al.~\cite{DBLP:journals/corr/abs-2209-02299}     & $\checkmark$ &  $\checkmark$            &  $\times$       & $\times$      &  $\checkmark$             & $\checkmark$          & $\checkmark$               &  $\checkmark$      &  $\checkmark$     & $\checkmark$        &  $\checkmark$        & $\checkmark$     &   $\times$     &   $\times$                        &   $\checkmark$          & $\checkmark$               & $\checkmark$                   &  $\times$            &     $\times$                    &     $\times$         &  $\times$        &  $\checkmark$            &   $\times$          \\ 

Saurabh et al.~\cite{DBLP:conf/apf/ShintreRD19}     & $\times$  &  $\times$            &  $\times$       & $\times$      &  $\times$             & $\times$          & $\times$               & $\checkmark$      &  $\checkmark$     & $\times$        &  $\times$        & $\times$      &  $\times$     &   $\times$                          &   $\times$          & $\times$              & $\times$                 &  $\times$            &   $\times$                 &     $\times$        &  $\times$        &  $\times$            &   $\times$          \\ 

\rowcolor{gray!40}
 Anvith et al.~\cite{DBLP:conf/eurosp/ThudiDCP22}     & $\checkmark$  &  $\checkmark$            &  $\times$        & $\times$       &  $\checkmark$             & $\times$          & $\times$                &  $\times$       &  $\times$      & $\times$         &  $\times$         & $\times$       &   $\times$    &   $\times$                          &   $\checkmark$         & $\times$                & $\times$                   &  $\times$            &   $\checkmark$                     &     $\times$          &  $\times$        &  $\times$           &  $\times$            \\ 

Ours     & $\checkmark$  &  $\checkmark$            &  $\checkmark$       & $\checkmark$      &  $\checkmark$             & $\checkmark$          & $\checkmark$               &  $\checkmark$      &  $\checkmark$     & $\checkmark$        &  $\checkmark$        & $\checkmark$      &   $\checkmark$     &   $\checkmark$               & $\checkmark$               &   $\checkmark$          & $\checkmark$               & $\checkmark$                   &  $\checkmark$            &     $\checkmark$                    &     $\checkmark$         &  $\checkmark$        &  $\checkmark$                  \\ \hline
\end{tabular}
}
\end{table}

\begin{itemize}
    \item Thanh et al.~\cite{DBLP:journals/corr/abs-2209-02299} summarized the definitions of machine unlearning, the unlearning request types, and different designing requirements. They also provided a taxonomy of the existing unlearning schemes based on available models and data. 
    \item Saurabh et al.~\cite{DBLP:conf/apf/ShintreRD19} analyzed the problem of privacy leakage in machine learning and briefly described how the “right-to-be-forgotten” can be implemented with the potential approaches.
    \item Anvith et al.~\cite{DBLP:conf/eurosp/ThudiDCP22} discussed the semantics behind unlearning and reviewed existing unlearning schemes based on logits, weights, and weight distributions. They also briefly described partial validation schemes of machine unlearning.
\end{itemize}

In addition to the difference in Table~\ref{tab:comparison}, this survey also differs from the above references in several aspects. Firstly, we provide a comprehensive analysis of each unlearning scheme together with corresponding verification strategies since the verification problem is an important metric in future studies. This is the significant difference between the above reference, as existing works have only reviewed the unlearning schemes used in each work. Second, each unlearning scheme is reviewed and compared through several dimensions, such as whether original training data is required, whether intermediate data needs to be cached, which classes and models are supported for unlearning requests, etc. In addition, we also analyze the commonalities and problems within each category in our taxonomy scheme, summarizing the trends, shortcomings and potential solutions, which have not been fully discussed in the above works~\cite{DBLP:journals/corr/abs-2209-02299,DBLP:conf/apf/ShintreRD19,DBLP:conf/eurosp/ThudiDCP22}. 

Our work also involves multiple key areas of privacy preserving and optimization, covering topics of differential privacy, data masking, convex optimization, and so on. In contrast, existing surveys mainly focus on summarizing the methods employed in machine unlearning, ignoring the relationship between unlearning strategy and verification technique. The most similar work to ours is~\cite{DBLP:journals/corr/abs-2209-02299}, however, it elaborates more on the unlearning framework and its application scenario, while we particularly emphasize unlearning strategy and verification. Moreover, we explore the possible trends of machine unlearning and summarise the latest research progress and possible techniques involved, including universality and security, etc., and suggest several specific research directions. Those are also not provided in the above reference~\cite{DBLP:journals/corr/abs-2209-02299,DBLP:conf/apf/ShintreRD19,DBLP:conf/eurosp/ThudiDCP22}.

\section{Preliminaries}
\label{sec:Preliminaries}


\subsection{Definition of Machine Unlearning}

\begin{table}
  \caption{Notations.}
  \label{tab:Notations}
  \resizebox{\textwidth}{!}{
  \begin{tabular}{c|c||c|c}
    \toprule
    Notations &  Explanation & Notations & Explanation\\
    \midrule
     $\mathcal{X}$          &The instance space 	            &$\mathcal{Y}$ 	        &The label space \\
     $\mathcal{D}$          &The training dataset 	            &$\mathcal{D}_{r}$      &The remaining dataset \\
     $\mathcal{D}_{u}$      &The unlearning dataset             &$\mathbf{x}_{i}$	    &One sample in $\mathcal{D}$ \\
     $y_{i}$		        &The label of sample $\mathbf{x}_{i}$&$n$				    &The size of $\mathcal{D}$\\
     $\mathbf{x}_{i,j}$     &The $j$-th feature in $\mathbf{x}_{i}$   &$d$                    &The dimension of $\mathbf{x}_{i}$\\
     $\mathcal{A}(\cdot)$   &The learning process               &$\mathcal{U}(\cdot)$   &The unlearning process\\
     $\mathcal{R}(\cdot)$   &The retraining process	            &$\mathbf{w}$           &The parameters of learned model  \\
     $\mathbf{w}_u$	        &The parameters of unlearned model  &$\mathbf{w}_r$	        &The parameters of retrained model\\
     $P(\cdot)$             &The distribution function          &$\mathcal{K}(\cdot)$   &The distribution measurement\\
     $\mathcal{I}(\cdot)$   &The shannon mutual information     &$\mathcal{H}$          &The hypothesis space for $\mathbf{w}$\\
    \bottomrule	
	\end{tabular}
	}
\end{table}

Vectors are denoted as bold lowercase, e.g., $\mathbf{x}_i$, and space or set as italics in uppercase, e.g., $\mathcal{X}$. A general definition of machine learning is given based on a supervised learning setting. The instance space is defined as  $\mathcal{X} \subseteq \mathbb{R}^{d}$, with the label space defined as $\mathcal{Y} \subseteq \mathbb{R}$. $\mathcal{D}=\left\{\left(\mathbf{x}_{i}, y_{i}\right)\right\}_{i=1}^{n} \subseteq \mathbb{R}^{d} \times \mathbb{R} $ represents a training dataset, in which each sample $\mathbf{x}_{i} \in \mathcal{X}$ is a $d$-dimensional vector $\left(x_{i, j}\right)_{j=1}^{d}$, $y_{i} \in \mathcal{Y}$ is the corresponding label, and $n$ is the size of $\mathcal{D}$. Let $d$ be the dimension of $\mathbf{x}_{i}$ and let $\mathbf{x}_{i,j}$ denote the $j$-th feature in the sample $\mathbf{x}_{i}$. 

The purpose of machine learning is to build a model $M$ with the parameters $\mathbf{w} \in \mathcal{H}$ based on a specific training algorithm $\mathcal{A}(\cdot)$, where $\mathcal{H}$ is the hypothesis space for $\mathbf{w}$. In machine unlearning, let $\mathcal{D}_{u} \subset \mathcal{D}$ be a subset of the training dataset, whose influence we want to remove from the trained model. Let its complement $\mathcal{D}_{r}=\mathcal{D}_{u}^{\complement} = \mathcal{D}/\mathcal{D}_{u}$ be the dataset that we want to retain, and let $\mathcal{R}(\cdot)$ and $\mathcal{U}(\cdot)$ represent the retraining process and unlearning process, respectively. $\mathbf{w}_{r}$ and $\mathbf{w}_{u}$ donate the parameters of the built models from those two processes. $P(a)$ represents the distribution of a variable $a$ and $\mathcal{K}(\cdot)$ represents a measurement of the similarity of two distributions. When considering $\mathcal{K}(\cdot)$ as a Kullback-Leibler (KL) divergence, $\mathcal{K}(\cdot)$ is defined by $\mathrm{KL}(P(a) \| P(b)):=\mathbb{E}_{a \sim P(a)}[\log (P(a) / P(b))]$. Given two random variables $a$ and $b$, the amount of Shannon Mutual Information that $a$ has about $b$ is defined as $I(a ; b)$. The main notations are summarized in Table ~\ref{tab:Notations}.

Now we give the definition of machine unlearning.

\begin{definition}[Machine Unlearning~\cite{DBLP:conf/sp/CaoY15}]
	\label{Definition:Machineunlearning}
	Consider a cluster of samples that we want to remove from the training dataset and the trained model, denoted as $\mathcal{D}_{u}$. An unlearning process $\mathcal{U}(\mathcal{A}(\mathcal{D}), \mathcal{D}, \mathcal{D}_{u})$ is defined as a function from an trained model $\mathcal{A}(\mathcal{D})$, a training dataset $\mathcal{D}$, and an unlearning dataset $\mathcal{D}_{u}$ to a model $\mathbf{w}_u$, which ensures that the unlearned model $\mathbf{w}_{u}$ performs as though it had never seen the unlearning dataset $\mathcal{D}_{u}$.
\end{definition}

\begin{figure}
  \centering
  \includegraphics[width=0.85\linewidth]{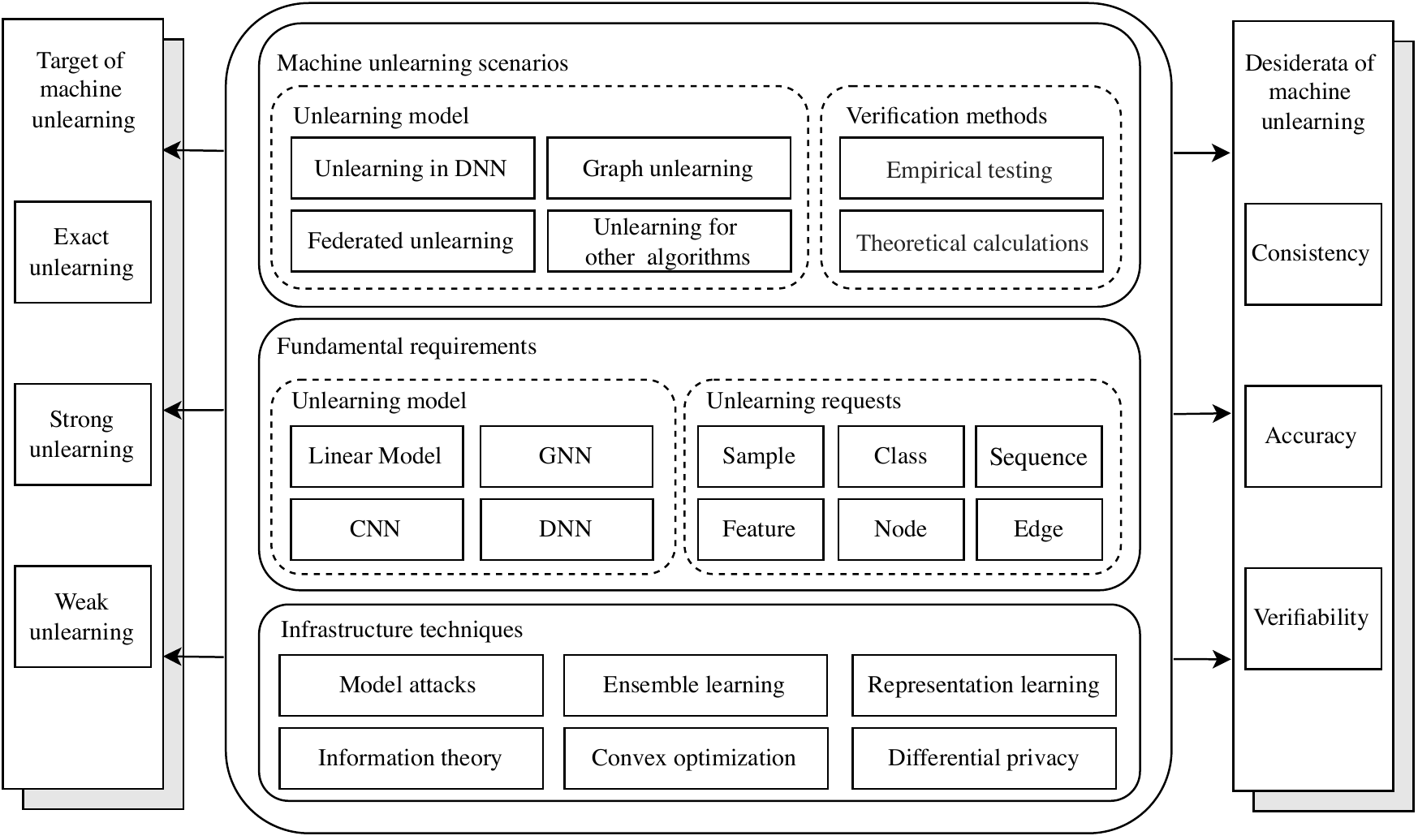}
  \caption{Machine Unlearning and Its Ecosystem.}
  \label{figure:insightview}
\end{figure}

Figure~\ref{figure:insightview} presents the typical concept, unlearning targets and desiderata associated with machine unlearning. The infrastructure techniques involved in machine unlearning include several aspects, such as ensemble learning, convex optimization, and so on~\cite{DBLP:conf/icml/KohL17}. These technologies provide robust guarantees for different foundational unlearning requirements that consists of various types of models and unlearning requests, resulting in diverse unlearning scenarios and corresponding verification methods. Additionally, to ensure effectiveness, the unlearning process requires different targets, such as exact unlearning or strong unlearning. Each unlearning target ensures different similarities in the distribution of the parameters between the unlearned model and that of the retrained model. Machine unlearning also involves several unlearning desiderata, including consistency, accuracy, and verifiability. Those desiderata, with the target constraint, simultaneously guarantee the validity and feasibility of each unlearning scheme.

\subsection{Targets of Machine Unlearning}
The ultimate target of machine unlearning is to reproduce a model that 1). behaves as if trained without seeing the unlearned data, and 2). consumes as less time as possible. The performance baseline of an unlearned model is that of the model retrained from scratch (a.k.a, native retraining).

\begin{definition}[Native retraining~\cite{DBLP:conf/sp/CaoY15}]
	\label{Definition:Nativeretraining} 
	Supposing the learning process, $\mathcal{A}(\cdot)$, never sees the unlearning dataset $\mathcal{D}_{u}$, and thereby performs a retraining process on the remaining dataset, denoted as $\mathcal{D}_{r} = \mathcal{D} \backslash \mathcal{D}_{u}$. In this manner, the retraining process is defined as:
	\begin{equation}
		\mathbf{w}_r = \mathcal{A}(\mathcal{D} \backslash \mathcal{D}_{u}) 
	\end{equation}
\end{definition}


The naive retraining naturally ensures that any information about samples can be unlearned from both the training dataset and the already-trained model.
However, the computational and time overhead associated with the retraining process could be significantly expensive. Further, a retraining process is not always possible if the training dataset is inaccessible, such as federated learning~\cite{DBLP:journals/corr/abs-2111-12056}. Therefore, two alternative unlearning targets have been proposed: exact unlearning and approximate unlearning. 

\begin{figure}
  \centering
  \includegraphics[width=0.8\linewidth]{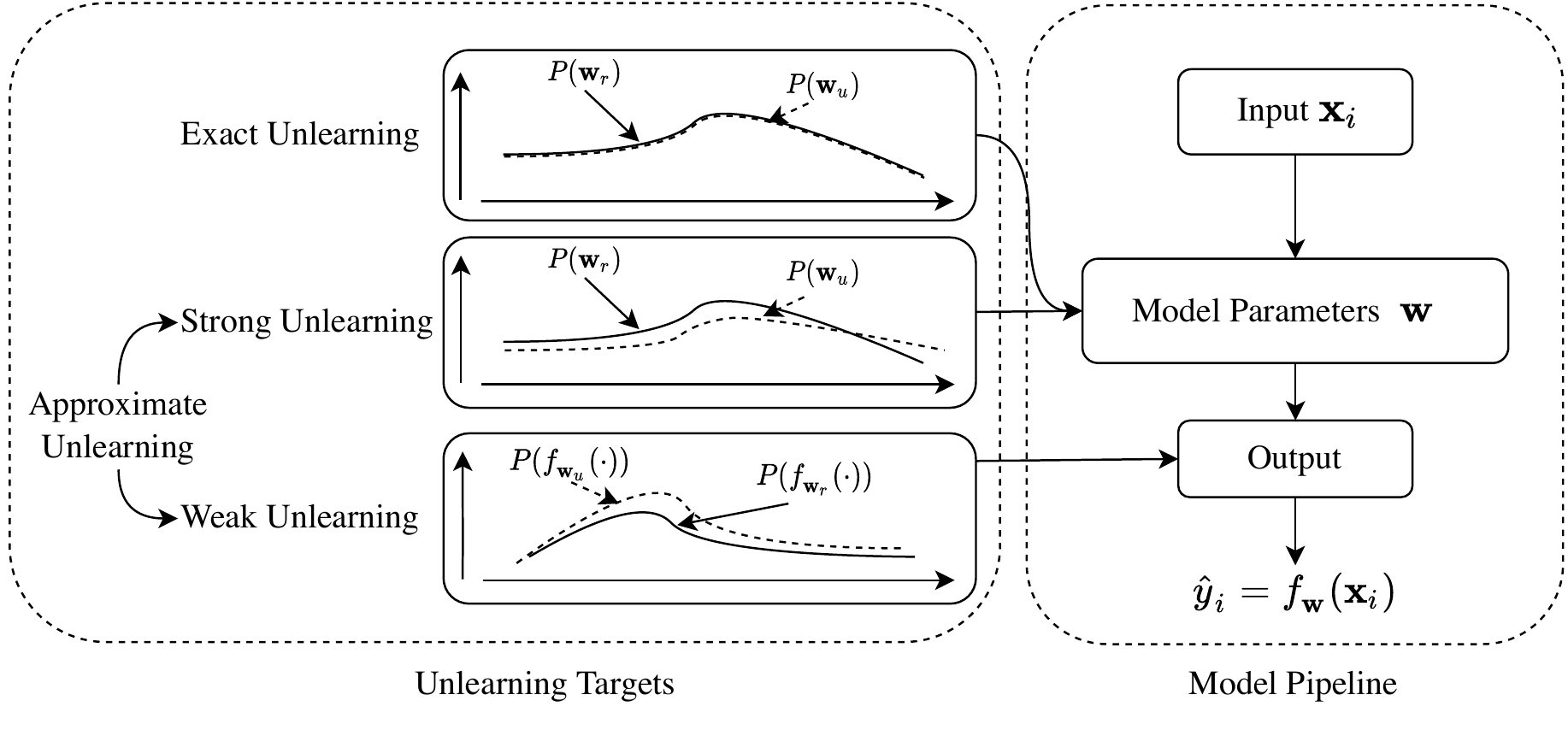}
  \caption{Targets of Machine Unlearning.}
  \label{figure:target_machine_learning}
\end{figure}

Exact unlearning guarantees that the distribution of an unlearned model and a retrained model are indistinguishable. In comparison, approximate unlearning mitigates the indistinguishability in weights and final activation, respectively. In practice, approximate unlearning further evolves to strong and weak unlearning strategies. Figure~\ref{figure:target_machine_learning} illustrates the targets of machine unlearning and their relationship with a trained model. The different targets are, in essence, correspond to the requirement of unlearning results.

\begin{definition}[Exact unlearning~\cite{DBLP:conf/cvpr/GolatkarAS20}]
	\label{Definition:Exactunlearning}
	Given a distribution measurement $\mathcal{K}(\cdot)$, such as KL-divergence, the unlearning process $\mathcal{U}(\cdot)$ will provide an \textit{exact unlearning} target if 
	\begin{equation}
		\mathcal{K}(P(\mathcal{U}(\mathcal{A}(\mathcal{D}), \mathcal{D}, \mathcal{D}_{u})),P(\mathcal{A}(\mathcal{D} \backslash \mathcal{D}_{u}))) = 0,
	\end{equation}
where $P(\cdot)$ denotes the distribution of the weights.
\end{definition}	

Exact unlearning guarantees the two output distributions are indistinguishable, and thus preventing an observer (e.g., attacker) to exact any information about $\mathcal{D}_{u}$. 

However, a less strict unlearning target is necessary because exact unlearning can only be achieved for simple and well-structured models~\cite{DBLP:conf/cidr/Schelter20}. As a result, approximate unlearning, which is suitable to complex machine learning models, is proposed.

\begin{definition}[Approximate unlearning~\cite{DBLP:conf/eurosp/ThudiDCP22}]
    If $\mathcal{K}(P(\mathcal{U}(\mathcal{A}(\mathcal{D}), \mathcal{D}, \mathcal{D}_{u})),P(\mathcal{A}(\mathcal{D} \backslash \mathcal{D}_{u})))$ is limited within a tolerable threshold, the unlearning process $\mathcal{U}(\cdot)$ is defined as strong unlearning.
\end{definition}


Approximate unlearning ensures that the distribution of the unlearned model and that of a retrained model are approximately indistinguishable. This approximation is usually guaranteed by differential privacy techniques, such as $(\varepsilon, \delta)$-certified unlearning~\cite{DBLP:conf/icml/GuoGHM20,DBLP:journals/corr/abs-2103-03279}.

Depending on where how the distribution is estimated, approximate unlearning can be further classified into \textit{strong unlearning} and \textit{weak unlearning}. Strong unlearning is established based on the similarity between the internal parameter distributions of the models, while weak unlearning is based on the distribution of the model's final activation results~\cite{DBLP:journals/corr/abs-2103-03279,DBLP:conf/eccv/GolatkarAS20}.


Table~\ref{tab:targetscomparison} summarizes the main differences between each unlearning target.

\begin{table}
    \caption{Summary and Comparison of Difference Between Targets.}
    \label{tab:targetscomparison}
	\resizebox{\textwidth}{!}{
	\begin{tabular}{llll}
	 \toprule	
	\begin{tabular}{l}Tartgets\end{tabular}
	&\begin{tabular}{l}Aims\end{tabular}	
	&\begin{tabular}{l}Advantages\end{tabular}  
	&\begin{tabular}{l}Limitations\end{tabular}\\
	
	\midrule
	\rowcolor{gray!40}
	\begin{tabular}{l}Exact Unlearning\end{tabular}&
	\begin{tabular}{l}To make the distributions\\
	                  of a natively retrained \\
	                  model and an unlearned \\
	                  model indistinguishable\end{tabular}&
	\begin{tabular}{l}Ensures that attackers \\
	                  cannot recover any \\
	                  information from the\\
	                  unlearned model\end{tabular}&
	\begin{tabular}{l}Difficult to implement\end{tabular} \\
	
	\midrule
	\begin{tabular}{l}Strong Unlearning\end{tabular}&
	\begin{tabular}{l}To ensure that the \\
	                  distributions of two \\
	                  models are approximately\\
	                  indistinguishable\end{tabular}&
	\begin{tabular}{l}Easier to implement \\
	                  than exact unlearning\end{tabular}&
	\begin{tabular}{l}Attackers can still\\
	                  recover some information\\
	                  from the unlearned model\end{tabular} \\
	
	\midrule
	\rowcolor{gray!40}
	\begin{tabular}{l}Weak Unlearning\end{tabular}&
	\begin{tabular}{l}To only ensure that \\
	                  the distributions of\\
	                  two final activations\\
	                  are indistinguishable\end{tabular}&
	\begin{tabular}{l}The easiest target for\\
	                  machine unlearning\end{tabular}&
	\begin{tabular}{l}Cannot guarantee \\
	                  whether the internal\\
	                  parameters of the model\\
	                  are successfully unlearned \end{tabular} \\

	\bottomrule
	\end{tabular}
	}
\end{table}

\subsection{Desiderata of Machine Unlearning}
\label{subsec:properties}

To fairly and accurately assess the efficiency and effectiveness of unlearning approaches, there are some mathematical properties that can be used for evaluation.

\begin{definition}[Consistency]
	Assume there is a set of samples $X_{e}$, with the true labels $Y_{e}:\left\{y_{1}^{e}, y_{2}^{e}, \ldots, y_{n}^{e}\right\}$. Let $Y_{n}:\left\{y_{1}^{n}, y_{2}^{n}, \ldots, y_{n}^{n}\right\}$ and $Y_{u}:\left\{y_{1}^{u}, y_{2}^{u}, \ldots, y_{n}^{u}\right\}$ be the predicted labels produced from a retrained model and an unlearned model, respectively. If all $y_{i}^{n}=y_{i}^{u}, 1 \leq i \leq n$, the unlearning process $\mathcal{U}(\mathcal{A}(\mathcal{D}), \mathcal{D},\mathcal{D}_{u})$ is considered to provide the consistency property.
\end{definition}

Consistency denotes how similar the behavior of a retrained model and an unlearned model is. It represents whether the unlearning strategy can effectively remove all the information of the unlearning dataset $\mathcal{D}_{u}$. If, for every sample, the unlearned model gives the same prediction result as the retrained model, then an attacker has no way to infer information about the unlearned data.

\begin{definition}[Accuracy]
	Given a set of samples $X_{e}$ in remaining dataset, where their true labels are $Y_{e}:\left\{y_{1}^{e}, y_{2}^{e}, \ldots, y_{n}^{e}\right\}$. Let $Y_{u}:\left\{y_{1}^{u}, y_{2}^{u}, \ldots, y_{n}^{u}\right\}$ to denote the predicted labels produced by the model after the unlearning process, $\mathbf{w}_{u} = \mathcal{U}(\mathcal{A}(\mathcal{D}), \mathcal{D},\mathcal{D}_{u})$. The unlearning process is considered to provide the accuracy property if all $y_{i}^{u}=y_{i}^{e}, 1 \leq i \leq n$.
\end{definition}

Accuracy refers to the ability of the unlearned model to predict samples correctly. It reveals the usability of a model after the unlearning process, given that a model with low accuracy is useless in practice. Accuracy is a key component of any unlearning mechanism, as we claim the unlearning mechanism is ineffective if the process significantly undermines the original model’s accuracy.
	
\begin{definition}[Verifiability]
	After the unlearning process, a verification function $\mathcal{V}(\cdot)$ can make a distinguishable check, that is, $\mathcal{V}\left(\mathcal{A}(\mathcal{D})\right) \neq \mathcal{V}\left(\mathcal{U}(\mathcal{A}(\mathcal{D}), \mathcal{D},\mathcal{D}_{u})\right)$.  The unlearning process $\mathcal{U}(\mathcal{A}(\mathcal{D}), \mathcal{D},\mathcal{D}_{u})$ can then provide a verifiability property. 
\end{definition}

Verifiability can be used to measure whether a model provider has successfully unlearned the requested unlearning dataset $\mathcal{D}_{u}$. Taking the following backdoor verification method as an example~\cite{DBLP:journals/corr/abs-2003-04247}, if the pre-injected backdoor for an unlearned sample $\mathbf{x}_{d}$ is verified as existing in $\mathcal{A}(\mathcal{D})$ but not $\mathcal{U}(\mathcal{A}(\mathcal{D}), \mathcal{D},\mathcal{D}_{u})$, that is $\mathcal{V}(\mathcal{A}(\mathcal{D})) = true$ and $\mathcal{V}(\mathcal{U}(\mathcal{A}(\mathcal{D}), \mathcal{D},\mathcal{D}_{u})) = false$, the unlearning method $\mathcal{U}(\mathcal{A}(\mathcal{D}), \mathcal{D},\mathcal{D}_{u})$ can be deemed to provide verifiability property.

\section{Taxonomy of Unlearning and Verification Mechanisms}
\label{sec:differentmethodologyinunlearningscheme}

\begin{figure}
  \centering
  \includegraphics[width=1\linewidth]{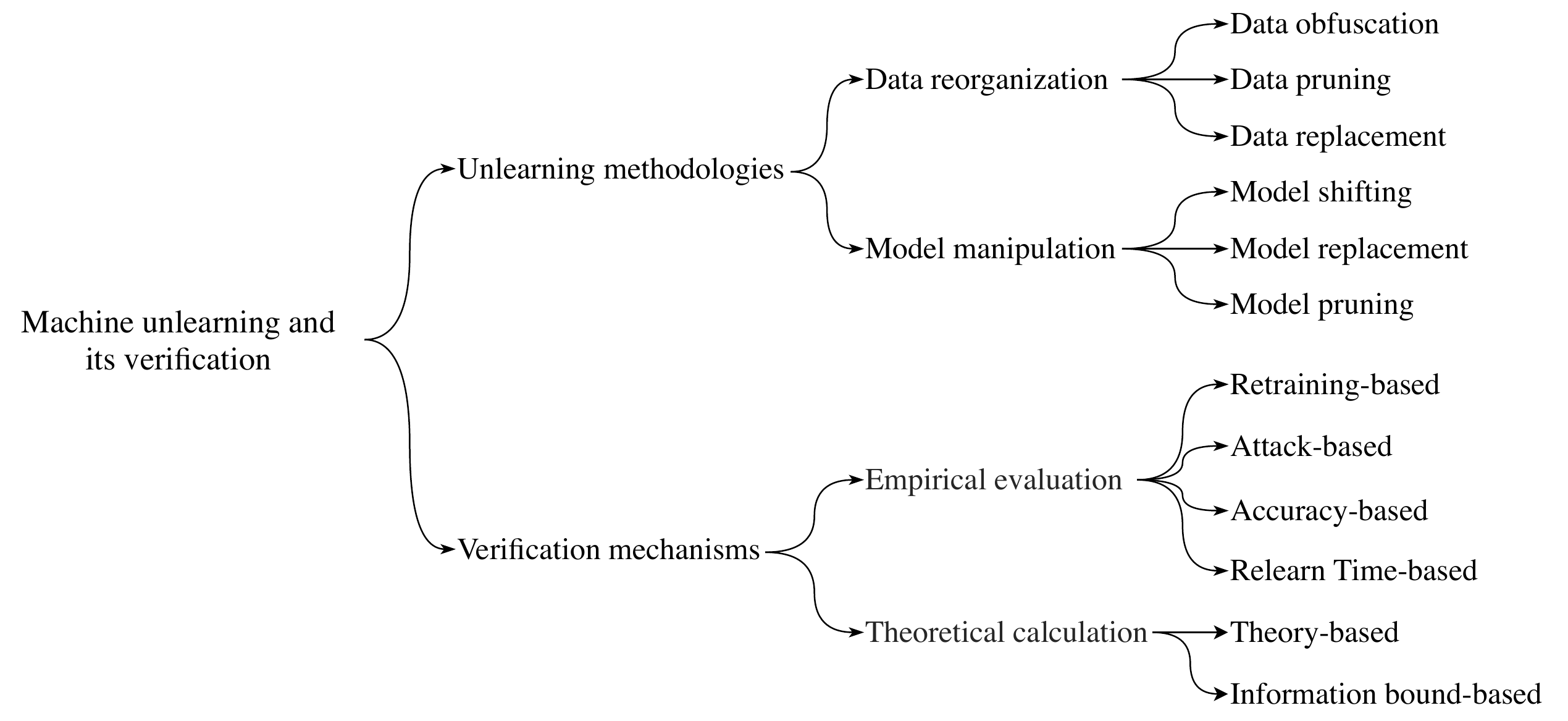}
  \caption{Taxonomy of Unlearning and Verification Mechanisms.}
  \label{figure:taxonomy}
\end{figure}

Figure~\ref{figure:taxonomy} summarizes the general taxonomy of machine unlearning and its verification used in this paper. The taxonomy is inspired by the design details of the unlearning strategy. Unlearning approaches that concentrate on modifying the training data are classified in data reorganization, while methods that directly manipulate the weights of a trained model are denoted as model manipulation. As for verification methods, initially, we categorize those schemes as either experimental or theoretical, subsequently, we summarize these methods based on the metrics they use.

\subsection{Unlearning Taxonomy}

\subsubsection{Data Reorganization}
\label{subsec:datareorganization}
    Data reorganization refers to the technique that a model provider unlearns data by reorganizing the training dataset. It mainly includes three different processing methods according to the different data reorganization modes: \textit{obfuscation}, \textit{pruning} and \textit{replacement}~\cite{DBLP:conf/sp/BourtouleCCJTZL21,DBLP:conf/aaai/GravesNG21}. Table~\ref{tab:classification} compares and summarizes the differences between these schemes.
 	\begin{itemize}
 	
        \item Data obfuscation: In data obfuscation, model providers intentionally add some choreographed data to the remaining dataset, that is $\mathcal{D}_{new} \leftarrow \mathcal{D}_{r} \cup \mathcal{D}_{obf}$, where $\mathcal{D}_{new}$ and $\mathcal{D}_{obf}$ are the new training dataset and the choreographed data, respectively. The trained model is then fine-tuned based on $\mathcal{D}_{new}$ to unlearn some specific samples. Such methods are usually based on the idea of erasing information about $\mathcal{D}_{u}$ by recombining the dataset with choreographed data. For example, Graves et al.~\cite{DBLP:conf/aaai/GravesNG21} relabeled $\mathcal{D}_{u}$ with randomly selected incorrect labels and then fine-tuned the trained model for several iterations for unlearning data.

        \item Data pruning: In data pruning, the model provider first segments the training dataset into several sub-datasets and trains several sub-models based on each sub-dataset. Those sub-models are then used to aggregate a consensus prediction collaboratively, that is $\mathcal{D} \rightarrow { \mathcal{D}_{1} \cup \mathcal{D}_{2} \cup ... \cup \mathcal{D}_{m}}$, $\mathbf{w}_{i} = \mathcal{A}(\mathcal{D}_{i})$ and $f(\mathbf{x}) = Agg(M_{\mathbf{w}_{i}}(\mathbf{x}))$, where $ \mathcal{D}_{i}$, $0 < i < m$ are the sub-datasets, and $\cap \mathcal{D}_{i}=\varnothing$, $\cup\mathcal{D}_{i}=\mathcal{D}$, $m$ is the number of sub-dataset, $\mathbf{w}_{i}$ is the sub-model, and $Agg(\cdot)$ is the aggregation function. After an unlearning request arrives, the model provider deletes the unlearned samples from the sub-datasets that contain them and then retrains the affected sub-models. The flexibility of this methodology is that the influence of unlearning dataset $\mathcal{D}_{u}$ is limited to each sub-dataset after segmentation rather than the whole dataset. Taking the $\mathbf{SISA}$ scheme in ~\cite{DBLP:conf/sp/BourtouleCCJTZL21} as an example, the $\mathbf{SISA}$ framework first randomly divided the training dataset into $k$ shards. A series of models are then trained separately at one per shard. When a sample needs to be unlearned, it is first removed from the shards that contain it, and only the sub-models corresponding to those shards are retrained.

        \item Data replacement: In data replacement, the model provider deliberately replaces the training dataset $\mathcal{D}$ with some new transformed dataset, that is $\mathcal{D}_{trans} \leftarrow \mathcal{D}$. The transformed dataset $\mathcal{D}_{trans}$ is then used to train a model that makes it easy to implement unlearning after receiving an unlearning request. For example, Cao et al.~\cite{DBLP:conf/sp/CaoY15} replaced the training dataset with several efficiently computable transformations and used those transformations to complete the training of the model. Those transformations can be updated much more quickly after removing any samples from the transformed dataset. Consequently, computational overheads are reduced, and unlearning operations are more efficient. 		
	 \end{itemize}

\newcolumntype{M}[1]{>{\centering\arraybackslash}m{#1}}
\begin{table}

    \caption{Summary and Comparison of Differences Between Unlearning Schemes.}
    \label{tab:classification}
    \resizebox{\textwidth}{!}{
    \begin{tabular}{M{0.1in}llll}
    \toprule
    \rowcolor{gray!40}
    &\begin{tabular}{l}Schemes\end{tabular}	                                                  				
    &\begin{tabular}{l}Basic Ideas\end{tabular}	
	&\begin{tabular}{l}Advantages\end{tabular}
	&\begin{tabular}{l}Limitations\end{tabular}\\
    \midrule

    \multirow{15}{*}{\rotatebox[origin=c]{90}{Data Reorganization}}

    &\begin{tabular}{l}
        Data \\ Obfuscation\\ 
	    \cite{DBLP:conf/aaai/GravesNG21,DBLP:conf/icores/FelpsSWVBHSSS21, DBLP:journals/corr/abs-2111-08947}
	    \end{tabular}
	    
    & \begin{tabular}{l}
	Intentionally adds \\
	some choreographed \\
	dataset to the \\
	training dataset and\\
	retrains the model\\
	\end{tabular}             
    
    &\begin{tabular}{l}
	Can be applied to \\
	almost all types of\\
	models; not too much\\
	intermediate redundant\\
	data need to be \\
	retained\end{tabular}

	&\begin{tabular}{l}
    Not easy to completely\\
    unlearn information\\
    from models\end{tabular}  \\
                                    
    &\cellcolor[HTML]{cbcbcb}\begin{tabular}{l}
    Data \\
    Pruning\\
	\cite{DBLP:conf/sp/CaoY15,DBLP:conf/sp/BourtouleCCJTZL21,DBLP:conf/ccs/Chen000H022}\\	\cite{DBLP:journals/corr/abs-2105-06209,Gupta2021adaptive,DBLP:conf/alt/Neel0S21}		\end{tabular}
	
	&\cellcolor[HTML]{cbcbcb}\begin{tabular}{l}
	Deletes the unlearned\\
	samples from sub-datasets\\
	that contains those \\
	unlearned samples.\\
	Then only retrains the \\
	sub-models that affected\\
	by those samples \end{tabular} 
           
    &\cellcolor[HTML]{cbcbcb}\begin{tabular}{l}
	Easy to implement and\\
	understand; completes\\
	the unlearning process\\
	at a faster speed\end{tabular}

	&\cellcolor[HTML]{cbcbcb}\begin{tabular}{l}
	Additional storage \\
	space is required; \\
	accuracy can be \\
	decreased with an\\
	increase in the \\
	number of sub-datasets\end{tabular}\\

    &\begin{tabular}{l}
	Data \\
	Replacement\\
	\cite{DBLP:conf/sp/CaoY15}\end{tabular}							
	
	&\begin{tabular}{l}
	Deliberately replaces\\
	the training dataset\\
	with some new \\
	transformed dataset\\
	\end{tabular} 
	
	&\begin{tabular}{l}
	Supports completely\\
	unlearn information\\
	from models; easy to\\
	implement\end{tabular}

	&\begin{tabular}{l}
	Hard to retain all\\
	the information about\\
	the original dataset\\
	through replacement \end{tabular} \\
    
    \midrule
    
    \multirow{12}{*}{\rotatebox[origin=c]{90}{Model Manipulation}}
    &\cellcolor[HTML]{cbcbcb}\begin{tabular}{l}
    Model \\
    Shifting\\ 
	~\cite{DBLP:conf/cidr/Schelter20, DBLP:conf/aaai/GravesNG21}	\\
	~\cite{DBLP:conf/icml/GuoGHM20,DBLP:conf/aistats/IzzoSCZ21,DBLP:conf/cvpr/GolatkarAS20}\\
	~\cite{DBLP:conf/eccv/GolatkarAS20,DBLP:conf/cvpr/GolatkarARPS21, DBLP:journals/corr/abs-2108-11577, DBLP:journals/corr/abs-2103-03279}\end{tabular}						

	&\cellcolor[HTML]{cbcbcb}\begin{tabular}{l}
	Directly updates \\
	model parameters \\
	to offset the impact\\
	of unlearned samples\\
	on the model\\
	\end{tabular} 
	
	&\cellcolor[HTML]{cbcbcb}\begin{tabular}{l}
	Does not require \\
	too much intermediate\\
	parameter storage;\\
    can provide theoretical\\
    verification
    \end{tabular}    
	
	&\cellcolor[HTML]{cbcbcb}\begin{tabular}{l}
	Not easy to find\\
	an appropriate offset\\
	value for complex\\
	models; calculating\\
	offset value is \\
	usually complex\\
	\end{tabular} \\
                                    
    & \begin{tabular}{l}
    Model \\
    Pruning\\ 
	~\cite{DBLP:conf/iwqos/LiuMYWL21, DBLP:conf/www/Wang0XQ22, DBLP:journals/corr/abs-2002-02730}\end{tabular}	
	
	&\begin{tabular}{l}
	Replaces partial\\
	parameters with \\
	pre-calculated  \\
	parameters\end{tabular}

	&\begin{tabular}{l}
	Reduces the cost \\
	caused by intermediate\\
	storage; the unlearning\\
	process can be completed\\
	at a faster speed\end{tabular}    
	
    &\begin{tabular}{l}
	Only applicable to\\
	partial models; not\\
	easy to implement \\
	and understand\end{tabular}  \\
                                    
    &\cellcolor[HTML]{cbcbcb}\begin{tabular}{l}
	Model\\ 
	Replacement\\ 
	~\cite{DBLP:conf/sigmod/SchelterGD21, DBLP:conf/icml/BrophyL21, DBLP:journals/corr/abs-2111-11869, DBLP:conf/icml/WuDD20}\end{tabular}	
	
	&\cellcolor[HTML]{cbcbcb}\begin{tabular}{l}
	Prunes some parameters \\
	from already-trained\\
	models
    \end{tabular}

	&\cellcolor[HTML]{cbcbcb}\begin{tabular}{l}
	Easy to completely\\
	unlearn information\\
	from models\end{tabular}

	&\cellcolor[HTML]{cbcbcb}\begin{tabular}{l}
	Only applicable to\\
	partial machine \\
	learning models; \\
	original model \\
	structure is usually\\
	changed\end{tabular}  \\
    
    \bottomrule
    \end{tabular}
    }
\end{table}

\subsubsection{Model Manipulation}
\label{subsec:modelmanipulation}
	In model manipulation, the model provider aims to realize unlearning operations by adjusting the model’s parameters. Model manipulation mainly includes the following three categories. Table~\ref{tab:classification} compares and summarizes the differences between these schemes.
	 \begin{itemize}
			\item Model shifting: In model shifting, the model providers directly update the model parameters to offset the impact of unlearned samples on the model, that is $\mathbf{w}_{u} = \mathbf{w} + \delta$, where $\mathbf{w}$ are parameters of the originally-trained model, and $\delta$ is the updated value. These methods are usually based on the idea of calculating the influence of samples on the model parameters and then updating the model parameters to remove that influence. It is usually extremely difficult to accurately calculate a sample’s influence on a model’s parameters, especially with complex deep neural models. Therefore, many model shifting-based unlearning schemes are based on specific assumptions. For example, Guo et al.’s~\cite{DBLP:conf/icml/GuoGHM20} unlearning algorithms are designed for linear models with strongly convex regularization.
			
			\item Model replacement: In model replacement, the model provider directly replaces some parameters with pre-calculated parameters, that is $\mathbf{w}_{u} \leftarrow \mathbf{w}_{noeffect} \cup \mathbf{w}_{pre}$, where $\mathbf{w}_{u}$ are parameters of the unlearned model, $\mathbf{w}_{noeffect}$ are partially unaffected static parameters, and $\mathbf{w}_{pre}$ are the pre-calculated parameters. These methods usually depend on a specific model structure to predict and calculate the affected parameters in advance. They are only suitable for some special machine learning models, such as decision trees or random forest models. Taking the method in~\cite{DBLP:conf/sigmod/SchelterGD21} as an example, the affected intermediate decision nodes are replaced based on pre-calculated decision nodes so as to generate an unlearned model.
			
			\item Model pruning: In model pruning, the model provider prunes some parameters from the trained models to unlearn the given samples, that is $\mathbf{w}_{u} \leftarrow \mathbf{w}/{\delta}$, where $\mathbf{w}_{u}$ are the parameters of the unlearned model, $\mathbf{w}$ are the parameters of the trained model, and $\delta$ are the parameters that need to be removed. Such unlearning schemes are also usually based on specific model structures and are generally accompanied by a fine-tuning process to recover performance after the model is pruned. For example, Wang et al.~\cite{DBLP:conf/www/Wang0XQ22} introduced the term frequency-inverse document frequency (TF-IDF) to quantize the class discrimination of channels in a convolutional neural network model, where channels with high TF-IDF scores are pruned.
	\end{itemize}

\begin{table}
    \caption{Summary and Comparison of Different Verification Methods.}
    \label{tab:verificationcomparison}
	\resizebox{\textwidth}{!}{
	\begin{tabular}{llll}
	 \toprule	
	\begin{tabular}{l}Methods\end{tabular}&\begin{tabular}{l}Basic Ideas\end{tabular}&\begin{tabular}{l}Advantages\end{tabular}&\begin{tabular}{l}Limitations\end{tabular}\\
	\midrule
	\rowcolor{gray!40}
	\begin{tabular}{l}Retraining-based\end{tabular}&
	\begin{tabular}{l}Removes unlearned\\
	                  samples and retrains\\
	                  models\end{tabular}&
	\begin{tabular}{l}Intuitive and\\
	                  easy to understand\end{tabular}&
	\begin{tabular}{l}Only applicable to\\
	                  special unlearning\\
	                  schemes\end{tabular}\\

	\begin{tabular}{l}Attack-based\end{tabular}&
	\begin{tabular}{l}Based on membership\\
	                  inference attacks or\\
	                  model inversion \\
	                  attacks\end{tabular}&
	\begin{tabular}{l}Intuitively measures\\
	                  the defense effect \\
	                  against some attacks\end{tabular}&
	\begin{tabular}{l}Inadequate verification\\
	                  capability\end{tabular}\\

 	\rowcolor{gray!40}
	\begin{tabular}{l}Relearn\\
	                  time-based\end{tabular}&
	\begin{tabular}{l}Measures the time when\\
	                  the unlearned model \\
	                  regains performance \\
	                  on unlearned samples\end{tabular}&
	\begin{tabular}{l}Easy to understand and\\
	                  easy to implement\end{tabular}&
	\begin{tabular}{l}Inadequate verification \\
	                  capability\end{tabular}\\

	\begin{tabular}{l}Accuracy-based\end{tabular}&
	\begin{tabular}{l}Same as a model trained\\
	                  without unlearned samples\end{tabular}&
	\begin{tabular}{l}Easy to understand and \\
	                  easy to implement\end{tabular}&
	\begin{tabular}{l}Inadequate verification \\
	                  capability\end{tabular}\\

 	\rowcolor{gray!40}
	\begin{tabular}{l}Theory-based\end{tabular}&
	\begin{tabular}{l}Ensures similarity \\
	                  between the unlearned\\
	                  model and the retrained\\
	                  model.\end{tabular}&
	\begin{tabular}{l}Comprehensive and \\
	                  has theoretical support\end{tabular}&
	\begin{tabular}{l}Implementation is\\
	                  complex and only \\
	                  applies to some \\
	                  specified models\end{tabular}\\  
	
	\begin{tabular}{l}Information \\bound-based\end{tabular}&
	\begin{tabular}{l}Measures the upper-bound\\
	                  of the residual\\
	                  information about the\\
	                  unlearned samples\end{tabular}&
	\begin{tabular}{l}Comprehensive and has\\
	                  theoretical support\end{tabular}&
	\begin{tabular}{l}Hard to implement and\\
	                  only applicable to some\\
	                  specified models\end{tabular}\\
	
	\bottomrule
	\end{tabular}
	}
\end{table}

\subsection{Verification Mechanisms}

Verifying whether the unlearning method has the verifiability property is not an easy task. Model providers may claim externally that they remove those influences from their models, but, in reality, this is not the case~\cite{DBLP:journals/corr/abs-2105-06209}. For data providers, proving that the model provider has completed the unlearning process may also be tricky, especially for complex deep models with huge training datasets. Removing a small portion of samples only causes a negligible effect on the model. Moreover, even if the unlearned samples have indeed been removed, the model still has a great chance of making a correct prediction since other users may have provided similar samples. Therefore, providing a reasonable unlearning verification mechanism is a topic worthy of further research.

\subsubsection{Empirical evaluation}
\begin{itemize}
	\item \textbf{Retraining-based verification}: Retraining can naturally provide a verifiability property, since the retraining dataset no longer contains the samples that need to be unlearned. This is the most intuitive and easy-to-understand solution.
	\item \textbf{Attack-based verification}: The essential purpose of an unlearning operation is to reduce leaks of sensitive information caused by model over-fitting. Hence, some attack methods can directly and effectively verify unlearning operations – for example, membership inference attacks~\cite{DBLP:conf/sp/ShokriSSS17} and model inversion attacks~\cite{DBLP:journals/tifs/KhosravyNHNB22}. In addition, Sommer et al.~\cite{DBLP:journals/corr/abs-2003-04247} provided a novel backdoor verification mechanism from an individual user perspective in the context of machine learning as a service (MLaaS)~\cite{DBLP:conf/mm/ZhangLH0YG20}. This approach can verify, with high confidence, whether the service provider complies with the user’s right to unlearn information.
	\item \textbf{Relearning time-based verification}: Relearning time can be used to measure the amount of information remaining in the model about the unlearned samples. If the model quickly recovers performance as the original trained model with little retraining time, then it is likely to still remember some information about the unlearned samples~\cite{DBLP:journals/corr/abs-2111-08947}.
	\item \textbf{Accuracy-based verification}: A trained model usually has high prediction accuracy for the samples in the training dataset. This means the unlearning process can be verified by the accuracy of a model’s output. For the data that need to be unlearned, the accuracy should ideally be the same as a model trained without seeing $\mathcal{D}_{u}$~\cite{DBLP:conf/cvpr/GolatkarAS20}. In addition, if a model’s accuracy after being attacked can be restored after unlearning the adversarial data, we can also claim that the unlearning is verified.

\end{itemize}

\subsubsection{Theoretical calculation}
\begin{itemize}
    	\item \textbf{Theory-based verification}: Some methods provide a certified unlearning definition~\cite{DBLP:conf/icml/GuoGHM20, DBLP:journals/corr/abs-2108-11577}, which ensures that the unlearned model cannot be distinguished from a model trained on the remaining dataset from scratch. This could also provide a verification method that directly guarantees the proposed schemes can unlearn samples.
	\item \textbf{Information bound-based verification}: Golatkar et al.~\cite{DBLP:conf/cvpr/GolatkarAS20,DBLP:conf/eccv/GolatkarAS20}, devised a new metric for verifying the effectiveness of unlearning schemes, where they measured the upper bound of the residual information about samples that need to be unlearned. Less residual information represents a more effective unlearning operation.
\end{itemize}

Table~\ref{tab:verificationcomparison} summarizes and compares each verification method's advantages and limitations.


\section{Data Reorganization}
\label{sec:datareorganization}
In this section, we review how data reorganization methods support the unlearning process. Since proving the verifiability property of unlearning algorithms is also important and should be considered in machine unlearning research, we separately discuss it for each unlearning method.

\subsection{Reorganization Based on Data Obfuscation}
\label{subsec:dataobfuscation}
\subsubsection{Unlearning Schemes Based on Data Obfuscation}~

\begin{figure}
  \centering
  \includegraphics[width=0.7\linewidth]{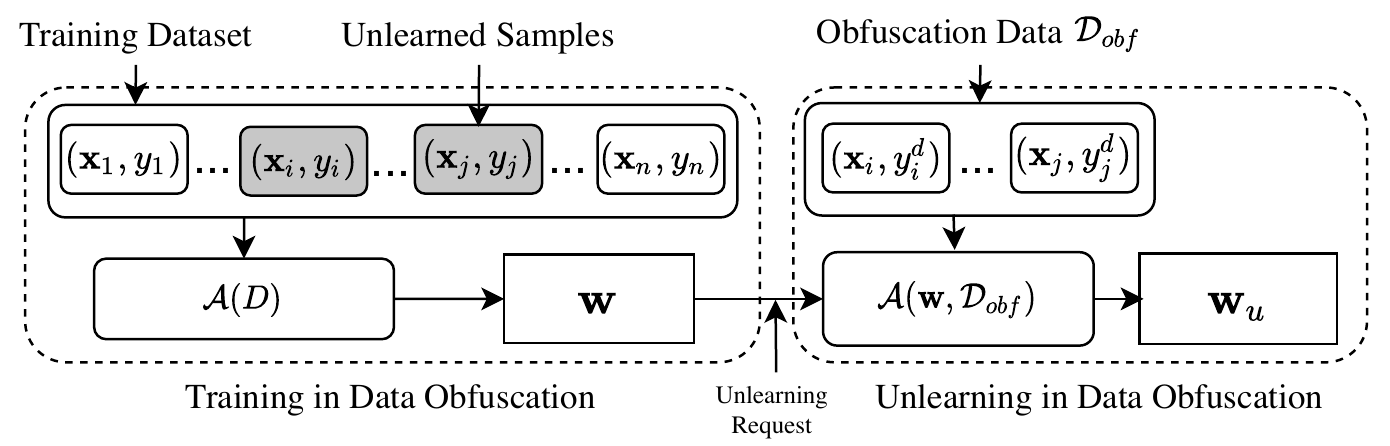}
  \caption{Unlearning Schemes Based on Data Obfuscation.}
  \label{fig:dataobfuscation}
\end{figure}

In general, the majority of model attack scenarios, such as membership inference attacks, arise from model overfitting and rely on observing shifts in the output based on known input shifts~\cite{DBLP:conf/ccs/Hu021}. That is, for the vast majority of attackers, it is easy to perform an attack on some trained models by observing the shifts of the output confidence vectors. One optional machine unlearning scheme can be interpreted as confusing the model’s understanding of samples so that it cannot retain any correct information within models. This method can further confuse the confidence vector of the model’s output~\cite{DBLP:conf/icores/FelpsSWVBHSSS21}. As shown in Figure~\ref{fig:dataobfuscation}, when receiving an unlearning request, the model continues to train $\mathbf{w}$ based on the constructed obfuscation data $\mathcal{D}_{obf}$ giving rise to an updated $\mathbf{w}_{u}$.

In this vein, Graves et al.~\cite{DBLP:conf/aaai/GravesNG21} proposed a random \textit{relabel} and \textit{retraining} machine unlearning framework. Sensitive samples are relabeled with randomly-selected incorrect labels, and then the machine learning model is fine-tuned based on the modified dataset for several iterations to unlearn those specific samples. Similarly, Felps et al.~\cite{DBLP:conf/icores/FelpsSWVBHSSS21} intentionally poisoned the labels of the unlearning dataset and then fine-tuned the model based on the new poisoned dataset. However, such unlearning schemes only confuse the relationship between the model outputs and the samples; the model parameters may still contain information about each sample.

The trained model is always trained by minimizing the loss for all classes. If one can learn a kind of noise that only maximizes the loss for some classes, those classes can be unlearned. Based on this idea, Tarrun et al.~\cite{DBLP:journals/corr/abs-2111-08947} divided the unlearning process into two steps, \textit{impair} and \textit{repair}. In the first step, an error-maximizing noise matrix is learned that consists of highly influential samples corresponding to the unlearning class. The effect of the noise matrix is somehow the opposite of the unlearning data, and can destroy the information of unlearned data to unlearn single/multiple classes. To repair the performance degradation caused by the model unlearning process, the \textit{repair} step further adjusted the model based on the remaining data. 

Similarly, Zhang et al.~\cite{DBLP:conf/mm/ZhangBHX22} considered the unlearning request in the image retrieval field. The approach developed involves creating noisy data using a generative method to adjust the weights of the retrieval model and achieve the unlearning purposes. They also proposed a new learning framework, which includes both static and dynamic learning branches, ensuring that the generated noisy data only affects the unlearning data being forgotten without affecting the contribution of other remaining data. However, the above two scheme consumes more time to generate noise for unlearning process, which will affect the efficiency of the unlearning process~\cite{DBLP:journals/corr/abs-2111-08947,DBLP:conf/mm/ZhangBHX22}.

\subsubsection{Verifiability of Schemes Based on Data Obfuscation}~

To verify their unlearning process, Graves et al.~\cite{DBLP:conf/aaai/GravesNG21} used two state-of-the-art attack methods – a model inversion attack and a membership inference attack – to evaluate how much information was retained in the model parameters about specific samples after the unlearning process – in other words, how much information might be leaked after the unlearning process. Their model inversion attack is a modified version of the standard model inversion attack proposed by Fredrikson et al.~\cite{DBLP:conf/ccs/FredriksonJR15}. The three modifications include: adjusting the process function to every $n$ gradient descent steps; adding a small amount of noise to each feature before each inversion; and modifying the number of attack iterations performed. These adjustments allowed them to analyze complex models. For the membership inference attack, they used the method outlined by Yeom et al. in~\cite{DBLP:conf/csfw/YeomGFJ18}. Felps et al.’s verifiability analysis is also based on the membership inference attack~\cite{DBLP:conf/icores/FelpsSWVBHSSS21}.

In comparison, Tarrun et al.~\cite{DBLP:journals/corr/abs-2111-08947} evaluated the verifiability through several measurements. They first assessed relearning time by measuring the number of epochs for the unlearned model to reach the same accuracy as the originally-trained model. Then, the distance between the original model, the model after the unlearning process, and the retrained model are further evaluated.

\subsection{Reorganization Based on Data Pruning}
\label{subsec:datapruning}
\subsubsection{Unlearning Schemes Based on Data Pruning}~

\begin{figure}
  \centering
  \includegraphics[width=1\linewidth]{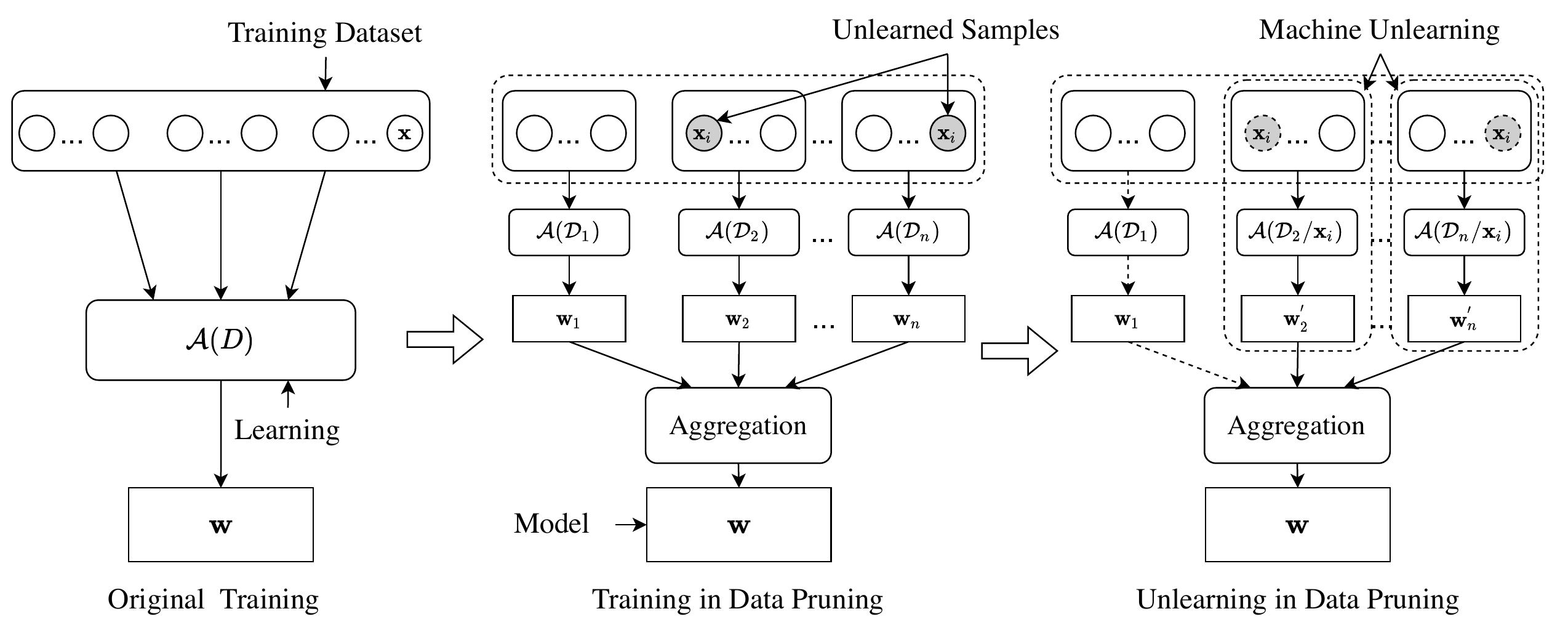}
  \caption{Unlearning Schemes Based on Data Pruning.}
  \label{fig:datapruning}
\end{figure}

As shown in Figure~\ref{fig:datapruning}, unlearning schemes based on data pruning are usually based on ensemble learning techniques. Bourtoule et al.~\cite{DBLP:conf/sp/BourtouleCCJTZL21} proposed a ``sharded, isolated, sliced, and aggregated'' ($\mathbf{SISA}$) framework, similar to the current distributed training strategies~\cite{DBLP:journals/jsac/HuynhHND22,DBLP:journals/tifs/GrattonVAW22}, as a method of machine unlearning. With this approach, the training dataset $\mathcal{D}$ is first partitioned into $k$ disjoint shards $\mathcal{D}_{1}, \mathcal{D}_{2}, \cdots, \mathcal{D}_{k}$. Then, sub-models $\mathcal{M}_{w}^{1}, \mathcal{M}_{w}^{2}, \cdots, \mathcal{M}_{w}^{k}$ are trained in isolation on each of these shards, which limits the influence of the samples to sub-models that were trained on the shards containing those samples. At inference time, $k$ individual predictions from each sub-model are simply aggregated to provide a global prediction (e.g., with majority voting), similar to the case of machine learning ensembles~\cite{DBLP:journals/tcs/JuniorFLNB20}. When the model owner receives a request to unlearn a data sample, they just need to retrain the sub-models whose shards contain that sample.  

As the amount of unlearning data increases, $\mathbf{SISA}$ will cause degradation in model performance, making them only suitable for small-scale scenarios. The cost of these unlearning schemes is the time required to retrain the affected sub-models, which directly relates to the size of the shard. The smaller the shard, the lower the cost of the unlearning scheme. At the same time, there is less training dataset for each sub-model, which will indirectly degrade the ensemble model’s accuracy. Bourtoule et al.~\cite{DBLP:conf/sp/BourtouleCCJTZL21} provided three key technologies to alleviate this problem, including \textit{unlearning in the absence of isolation}, \textit{data replication}, and \textit{core-set selection}.

In addition to this scheme, Chen et al.~\cite{DBLP:conf/www/Chen0ZD22} introduced the method developed in ~\cite{DBLP:conf/sp/BourtouleCCJTZL21} to recommendation systems and designed three novel data partition algorithms to divide the recommendation training data into balanced groups in order to ensure that collaborative information was retained. Wei et al.~\cite{DBLP:conf/bibm/QianZSCWH22} focused on the unlearning problems in patient similarity learning and proposed \textit{PatEraser}. To maintain the comparison information between patients, they developed a new data partition strategy that groups patients with similar characteristics into multiple shards. Additionally, they also proposed a novel aggregation strategy to improve the global model utility. 

Yan et al.~\cite{DBLP:conf/ijcai/YanLG0L022} designed an efficient architecture for exact machine unlearning, called \textit{ARCANE}, similar as the scheme in Bourtoule et al.~\cite{DBLP:conf/sp/BourtouleCCJTZL21}. Instead of dividing the dataset uniformly, they split it by class and utilized the one-class classifier to reduce the accuracy loss. Additionally, they also preprocessed each sub-dataset to speed up model retraining, which involved representative data selection, model training state saving, and data sorting by erasure probability. Nevertheless, the above unlearning schemes~\cite{DBLP:conf/sp/BourtouleCCJTZL21,DBLP:conf/www/Chen0ZD22,DBLP:conf/ijcai/YanLG0L022} usually need to cache a large number of intermediate results to complete the unlearning process. This will consume a lot of storage space.

$\mathbf{SISA}$ is designed to analyze Euclidean space data, such as images and text, rather than non-Euclidean space data, such as graphs. By now, numerous important real-world datasets are represented in the form of graphs, such as social networks~\cite{DBLP:conf/aaai/HuangXXDXLBXLY21}, financial networks~\cite{DBLP:journals/pr/ChengYXL22}, biological networks~\cite{DBLP:conf/bcb/KesimogluB21}, or transportation networks~\cite{DBLP:conf/aaai/Diao0ZLXH19}. To analyze the rich information in these graphs, graph neural networks (GNNs) have shown unprecedented advantages~\cite{DBLP:conf/ijcai/LambGGPAV20,DBLP:journals/tnn/WuPCLZY21}. GNNs rely on the graph’s structural information and neighboring node features. Yet naively applying $\mathbf{SISA}$ scheme to GNNs for unlearning, i.e., randomly partitioning the training dataset into multiple sub-graphs, will destroy the training graph’s structure and may severely damage the model’s utility.

To allow efficient retraining while keeping the structural information of the graph dataset, Chen et al.~\cite{DBLP:conf/ccs/Chen000H022} proposed \textit{GraphEraser}, a novel machine unlearning scheme tailored to graph data. They first defined two common machine unlearning requests in graph scenario: node unlearning and edge unlearning, and proposed a general pipeline for graph unlearning, which is composed of three main steps: \textit{graph partitioning}, \textit{shard model training}, and \textit{shard model aggravation}. In the \textit{graph partitioning} step, they introduced an improved balanced \textit{label propagation algorithm} (LPA)~\cite{DBLP:journals/isci/HeWCDCFWH21} and a balanced \textit{embedding $k$-means}~\cite{DBLP:conf/aaai/RenS021} partitioning strategy to avoid highly unbalanced shard sizes. Given that the different sub-models might provide different contributions to the final prediction, they also proposed a learning-based aggregation method, \textit{OptAggr}, that optimizes the importance score of each sub-model to improve global model utility ultimately.

Deterministic unlearning schemes, such as $\mathbf{SISA}$~\cite{DBLP:conf/sp/BourtouleCCJTZL21} or \textit{GraphEraser}~\cite{DBLP:conf/ccs/Chen000H022}, promise nothing about what can be learned about specific samples from the difference between a trained model and an unlearned model. This could exacerbate user privacy issues if an attacker has access to the model before and after the unlearning operation~\cite{DBLP:conf/ccs/Chen000HZ21}. To avoid this situation, an effective approach is to hide the information about the unlearned model when performing the unlearning operation.

In practical applications, Neel et al.~\cite{DBLP:conf/alt/Neel0S21} proposed an update-based unlearning method that performs several gradient descent updates to build an unlearned model. The method is designed to handle arbitrarily long sequences of unlearning requests with stable run-time and steady-state errors. In addition, to alleviate the above unlearning problem, they introduced the concept of \textit{secret state}: an unlearning operation is first performed on the trained model. Then, the unlearned models are perturbed by adding Gaussian noise for publication. This effectively ensures that an attacker cannot access the unlearned model actually after the unlearning operation, which effectively hides any sensitive information in the unlearned model. They also provided an $(\epsilon, \delta)$-certified unlearning guarantee, and leveraged a distributed optimization algorithm and reservoir sampling to grant improved accuracy/run-time tradeoffs for sufficiently high dimensional data.

After the initial model deployment, data providers may make an adaptive unlearning decision. For example, when a security researcher releases a new model attack method that identifies a specific subset of the training dataset, the owners of these subsets may rapidly increase the number of deletion requests. Gupta et al. ~\cite{Gupta2021adaptive} define the above unlearning requests as adaptive requests and propose an adaptive sequential machine unlearning method using a variant of the $\mathbf{SISA}$ framework~\cite{DBLP:conf/sp/BourtouleCCJTZL21} as well as a differentially private aggregation method~\cite{DBLP:journals/corr/abs-1803-10266}. They give a general reduction of the unlearning guarantees from the adaptive sequences to the non-adaptive sequences using differential privacy and max-information theory~\cite{DBLP:conf/stoc/BrownBFST21}. A strong provable unlearning guarantee for adaptive unlearning sequences is also provided, combined with the previous works of non-adaptive guarantees for sequence unlearning requests.

He et al.~\cite{DBLP:journals/corr/abs-2105-06209} developed an unlearning approach for the deep learning model. They first introduce a process called \textit{detrended fluctuation analysis}~\cite{PhysRevE.49.1685}, which quantifies the influence of the unlearned data on the model parameters, termed \textit{temporal residual memory}. They observed that this influence is subject to exponential decay, which fades at an increasing rate over time. Based on these results, intermediate models are retained during the training process and divided into four areas, named \textit{unseen}, \textit{deleted}, \textit{affected} and \textit{unaffected}. \textit{Unseen} indicates that the unlearned sample has not yet arrived. \textit{Deleted} includes the unlearning dataset. \textit{Unaffected} and \textit{affected} indicate whether temporal residual memory has lapsed or not. An unlearned model can be stitched by reusing the \textit{unseen} and \textit{unaffected} models and retraining the \textit{affected} areas. However, this scheme does not provide any theoretical verification methods to ensure that the information about unlearning data to be unlearned is indeed removed from the model.

\subsubsection{Verifiability of Schemes Based on Data Pruning}~

The unlearning schemes proposed in ~\cite{DBLP:conf/sp/CaoY15,DBLP:conf/sp/BourtouleCCJTZL21,DBLP:conf/www/Chen0ZD22,DBLP:conf/bibm/QianZSCWH22,DBLP:conf/ijcai/YanLG0L022,DBLP:conf/ccs/Chen000H022} are essentially based on a retraining mechanism that naturally has a verifiability property. As discussed in Section~\ref{subsec:properties}, a straightforward way to give an unlearning scheme the verifiability property is to retrain the model from scratch after removing the samples that need to be unlearned from the training dataset. The above schemes introduce distributed and ensemble learning techniques, which train sub-models separately and independently to optimize the loss function on each sub-dataset. The sub-models are then aggregated to make predictions. In terms of the unlearning process, only the affected sub-models are retrained, which avoids a large computational and time overhead and also provides a verifiability guarantee.

He et al.~\cite{DBLP:journals/corr/abs-2105-06209} use a backdoor verification method in~\cite{DBLP:journals/corr/abs-2003-04247} to verify their unlearning process. They designed a specially-crafted trigger and implanted this “backdoor data” in the samples that need to be unlearned, with little effect on the model’s accuracy. They indirectly verify the validity of the unlearning process based on whether the backdoor data can be used to attack the unlearned model with a high success rate. If the attack result has lower accuracy, it proves that the proposed unlearning method has removed the unlearned data. The other studies~\cite{DBLP:conf/alt/Neel0S21, Gupta2021adaptive} did not provide a method for verifying the unlearning process.

\subsection{Reorganization Based on Data Replacement}
\label{subsec:datareplacement}
\subsubsection{Unlearning Schemes Based on Data Replacement}~

\begin{figure}
  \centering
  \includegraphics[width=1.\linewidth]{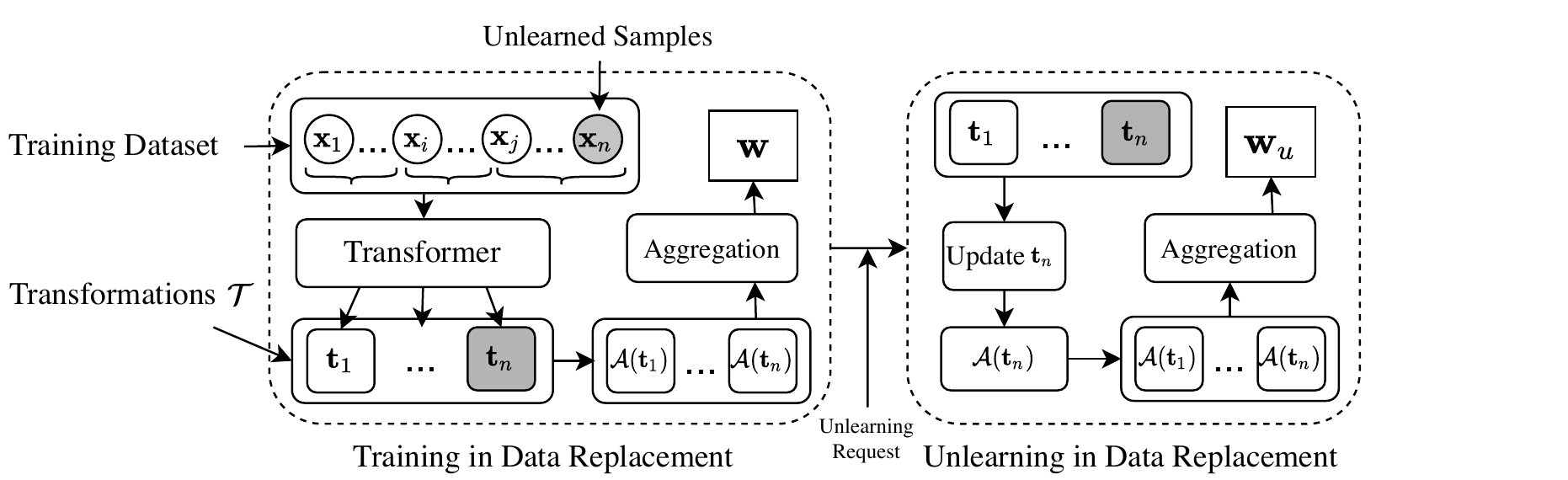}
  \caption{Unlearning Schemes Based on Data Replacement.}
  \label{fig:datareplacement}
\end{figure}

As shown in Figure~\ref{fig:datareplacement}, when training a model in a data replacement scheme, the first step is usually to transform the training dataset into an easily unlearned type, named transformation $\mathcal{T}$. Those transformations are then used to separately train models. When an unlearning request arrives, only a portion of the transformations $\mathbf{t}_i$ – the ones that contain the unlearned samples – need to be updated and used to retrain each sub-model to complete the machine unlearning.

Inspired by the previous work of using MapReduce to accelerate machine learning algorithms ~\cite{DBLP:journals/tcbb/SinhaPD22}, Cao et al.~\cite{DBLP:conf/sp/CaoY15} proposed a machine unlearning method that transforms the training dataset into summation form. Each summation is the sum of some efficiently computable transformation. The learning algorithms depend only on the summations, not the individual data, which breaks down the dependencies in the training dataset. To unlearn a data sample, the model provider only needs to update the summations affected by this sample and recompute the model. However, since the summation form comes from statistical query (SQ) learning, and only a few machine learning algorithms can be implemented as SQ learning, such as na\"ive bayes classifiers~\cite{DBLP:conf/aaai/ChoiFBB20}, support vector machines~\cite{DBLP:conf/icml/GronlundKL20}, and k-means clustering~\cite{DBLP:journals/pami/XiaPMZWGWC22}, this scheme has low applicability.

Takashi et al.~\cite{DBLP:conf/ijcai/0001IIM21} proposed a novel approach to lifelong learning named "Learning with Selective Forgetting", which involves updating a model for a new task by only forgetting specific classes from previous tasks while keeping the rest. To achieve this, the authors designed specific mnemonic codes, which are class-specific synthetic signals that are added to all the training samples of corresponding classes. Then, exploiting the mechanism of catastrophic forgetting, these codes were used to forget particular classes without requiring the original data. It is worth noting, however, that this scheme lacks any theoretical verification methods to confirm that the unlearning data information has been successfully removed from the model.

\subsubsection{Verifiability of Schemes Based on Data Replacement}~

Cao et al.~\cite{DBLP:conf/sp/CaoY15} provide an accuracy-based verification method. Specifically, they attack the LensKit model with the system inference attack method proposed by Calandrino et al.~\cite{ DBLP:conf/sp/CalandrinoKNFS11} and verify that the unlearning operations successfully prevent the attack from yielding any information. For the other three models, they first performed data pollution attacks to influence the accuracy of those models. They then analyzed whether the model’s performance after the unlearning process was restored to the same state as before the pollution attacks. If the unlearned model was actually restored to its pre-pollution value, the unlearning operation was considered to be successful. Takashi et al.~\cite{DBLP:conf/ijcai/0001IIM21} provided a new metric, named \textit{Learning with Selective Forgetting Measure (LSFM)} that is based on the idea of accuracy.

\setlength\rotFPtop{0pt plus 1fil} 
\begin{sidewaystable}
    \caption{The Surveyed Studies That Employed Data Reorganization Techniques for Unlearning Process.}
    \label{tab:reorganization}
	\resizebox{\textwidth}{!}{%
	\begin{tabular}{cccccccccc}
	 \toprule	
	Papers
	& \begin{tabular}{c}Unlearning \\Methods\end{tabular} 
	& \begin{tabular}{c}Unlearning \\Target\end{tabular} 
	& Training Dataset
	& Intermediates 
	& Unlearned Samples' Type                                         
	& Target Models' Type                        
	& Consistency 
	& Accuracy 
	& Verifiability\\
	
	\midrule
	
	\rowcolor{gray!40} 
	Graves et al.~\cite{DBLP:conf/aaai/GravesNG21} 
	& \begin{tabular}{c}Data \\Obfuscation\end{tabular} 
	& \begin{tabular}{c}Strong \\unlearning \end{tabular} 
	& Yes           
	& No            
	& Samples or Class                                     
	& DNN                               
	& No          
	& No       
	& Attack-Based\\
	
	Felps et al.~\cite{DBLP:conf/icores/FelpsSWVBHSSS21}
	& \begin{tabular}{c}Data \\Obfuscation\end{tabular} 
	& \begin{tabular}{c}Strong \\unlearning \end{tabular} 
	& No            
	& No            
	& Sequences                                            
	& DNN                               
	& No          
	& No       
	& Attack-Based                       \\
	
	\rowcolor{gray!40}
	Tarrun et al.~\cite{DBLP:journals/corr/abs-2111-08947} 
	& \begin{tabular}{c}Data \\Obfuscation\end{tabular} 
	& \begin{tabular}{c}Strong \\unlearning \end{tabular} 
	& Yes           
	& No            
	& Classes                                              
	& DNN                               
	& No          
	& No       
	& \begin{tabular}{c}Accuracy-Based \\ and\\Retrain Time-Based  \end{tabular} \\

    Zhang et al.~\cite{DBLP:conf/mm/ZhangBHX22}
	& \begin{tabular}{c}Data \\Obfuscation\end{tabular} 
	& \begin{tabular}{c}Strong \\unlearning \end{tabular} 
	& No           
	& No            
	& Samples                                              
	& DNN                               
	& No          
	& No       
	& \begin{tabular}{c}Accuracy-Based \\ and\\Retrain Time-Based  \end{tabular} \\
 
    \rowcolor{gray!40}  
	Bourtoule et al.~\cite{DBLP:conf/sp/BourtouleCCJTZL21}  
	& \begin{tabular}{c}Data \\Pruning\end{tabular} 
	& \begin{tabular}{c}Strong \\unlearning \end{tabular} 
	& Yes           
	& Yes           
	& \begin{tabular}{c}Batches \\and \\ Sequences\end{tabular} 
	& DNN                               
	& No          
	& No       
	& Retrain-Based   \\

	Chen et al.~\cite{DBLP:conf/www/Chen0ZD22} 
	& \begin{tabular}{c}Data \\Pruning\end{tabular} 
	& \begin{tabular}{c}Strong \\unlearning \end{tabular} 
	& Yes           
	& Yes           
	& \begin{tabular}{c}Samples\end{tabular} 
	& DNN                               
	& No          
	& No       
	& Retrain-Based   \\

    \rowcolor{gray!40}  
	Wei et al.~\cite{DBLP:conf/bibm/QianZSCWH22}
	& \begin{tabular}{c}Data \\Pruning\end{tabular} 
	& \begin{tabular}{c}Strong \\unlearning \end{tabular} 
	& Yes           
	& Yes           
	& \begin{tabular}{c}Samples\end{tabular} 
	& DNN                               
	& No          
	& No       
	& Retrain-Based   \\

    Yan et al.~\cite{DBLP:conf/ijcai/YanLG0L022}
	& \begin{tabular}{c}Data \\Pruning\end{tabular} 
	& \begin{tabular}{c}Exact \\unlearning \end{tabular} 
	& Yes           
	& Yes           
	& \begin{tabular}{c}Samples\end{tabular} 
	& DNN                               
	& Yes          
	& Yes      
	& Retrain-Based   \\
 
    \rowcolor{gray!40}  
	Chen et al.~\cite{DBLP:conf/ccs/Chen000H022} 
	& \begin{tabular}{c}Data \\Pruning\end{tabular} 
	& \begin{tabular}{c}Exact \\unlearning \end{tabular}
	& Yes           
	& Yes           
	& Nodes and Edges                                      
	& GNN                               
	& No          
	& No       
	& Retrain-Based \\ 

	Neel et al.~\cite{DBLP:conf/alt/Neel0S21}  
	& \begin{tabular}{c}Data \\Pruning\end{tabular} 
	& \begin{tabular}{c}Strong \\unlearning \end{tabular} 
	& Yes          
	& Yes           
	& \begin{tabular}{c} Non-adaptive\\Sequences \end{tabular}                               
	& Convex Model                      
	& No          
	& No       
	& Retrain-Based \\

    \rowcolor{gray!40}  
	Gupta et al.   ~\cite{Gupta2021adaptive}   
	& \begin{tabular}{c}Data \\Pruning\end{tabular} 
	& \begin{tabular}{c}Strong \\unlearning \end{tabular}
	& Yes           
	& Yes           
	& Adaptive Sequences                                   
	& Non-convex Models                 
	& No          
	& No       
	& Retrain-Based \\

	He et al.~\cite{DBLP:journals/corr/abs-2105-06209}  
	& \begin{tabular}{c}Data \\Pruning\end{tabular} 
	& \begin{tabular}{c}Strong \\unlearning \end{tabular} 
	& Yes           
	& Yes           
	& Samples                                              
	& DNN                               
	& No          
	& No       
	& Attack-Based  \\
	
 	\rowcolor{gray!40}
	Cao et al.~\cite{DBLP:conf/sp/CaoY15}  
	& \begin{tabular}{c}Data \\Replacement\end{tabular} 
	& \begin{tabular}{c}Exact \\Unlearning \end{tabular} 
	& Yes           
	& Yes          
	& Samples                                              
	&\begin{tabular}{c} Statistical Query \\Learning Models   \end{tabular} 
	& Yes         
	& Yes      
	& \begin{tabular}{c} Retrain-Based  \\ and\\ Accuracy-Based   \end{tabular}  \\

    Takashi et al.~\cite{DBLP:conf/ijcai/0001IIM21} 
	& \begin{tabular}{c}Data \\Replacement\end{tabular} 
	& \begin{tabular}{c}Strong \\unlearning \end{tabular} 
	& No           
	& Yes            
	& Classes                                              
	& DNN                               
	& No          
	& No       
	& \begin{tabular}{c}Accuracy-Based\end{tabular} \\
 
	\bottomrule
\end{tabular}
}
\end{sidewaystable}

\subsection{Summary of Data Reorganization}

In these last few subsections, we reviewed the studies that use data obfuscation, data pruning, and data replacement techniques as unlearning methods. A summary of the surveyed studies is shown in Table~\ref{tab:reorganization}, where we present the key differences between each paper.

From those summaries, we can see that most unlearning algorithms retain intermediate parameters and make use of the original training dataset~\cite{DBLP:conf/sp/BourtouleCCJTZL21,DBLP:conf/ccs/Chen000H022}. This is because those schemes usually segment the original training dataset and retrain the sub-models that were trained on the segments containing those unlearned samples. Consequently, the influence of specific samples is limited to only some of the sub-models and, in turn, the time taken to actually unlearn the samples is reduced. However, segmenting decreases time at the cost of additional storage. Thus, it would be well worth researching more efficient unlearning mechanisms that ensure the validity of the unlearning process and do not add too many storage costs simultaneously.

Moreover, these unlearning schemes usually support various unlearning requests and models, ranging from samples to classes or sequences and from support vector machines to complex deep neural models~\cite{DBLP:conf/sp/CaoY15,DBLP:conf/ccs/Chen000H022,DBLP:conf/alt/Neel0S21}. Unlearning schemes based on data reorganization rarely operate on the model directly. Instead, they achieve the unlearning purpose by modifying the distribution of the original training datasets and indirectly changing the obtained model. The benefit is that such techniques can be applied to more complex machine learning models. In addition to their high applicability, most of them can provide a strong unlearning guarantee, that is, the distribution of the unlearned model is approximately indistinguishable to that obtained by retraining.

It is worth pointing out that unlearning methods based on data reorganization will affect the consistency and the accuracy of the model as the unlearning process continues ~\cite{DBLP:conf/sp/BourtouleCCJTZL21,DBLP:conf/ccs/Chen000H022,DBLP:journals/corr/abs-2105-06209}. This reduction in accuracy stems from the fact that each sub-model is trained on the part of the dataset rather than the entire training dataset. This phenomenon does not guarantee that the accuracy of the unlearned model is the same as the result before the segmentation. Potential solutions are \textit{to use unlearning in the absence of isolation}, \textit{data replication}~\cite{DBLP:conf/sp/BourtouleCCJTZL21}.

Some of the studies mentioned indirectly verify the unlearning process using a retraining method~\cite{DBLP:conf/sp/BourtouleCCJTZL21,DBLP:conf/ccs/Chen000H022}, while others provide verifiability through attack-based or accuracy-based methods ~\cite{DBLP:conf/aaai/GravesNG21,DBLP:conf/icores/FelpsSWVBHSSS21,DBLP:journals/corr/abs-2111-08947}. However, most unlearning schemes do not present further investigations at the theoretical level. The vast majority of the above unlearning schemes verify validity through experiments, with no support for the theoretical validity of the schemes. Theoretical validity would show, for example, how much sensitive information attackers can glean from an unlearned model after unlearning process or how similar the parameters of the unlearned model are to the retrained model. Further theoretical research into the validity of unlearning schemes is therefore required.

In summary, when faced with unlearning requests for complex models, unlearning schemes based on data obfuscation seldom unlearn information. This is because it is difficult to offset the influence of the unlearning data completely. Data pruning schemes always affect the model's accuracy since they usually train sub-models using a partial training dataset.  For data replacement schemes, it is impossible to find a new dataset that can replace all the information within an original dataset to train a model. Thus, researchers should turn to design unlearning schemes that strike more of a balance between the effectiveness of the unlearning process and model usability.

\section{Model Manipulation}
\label{sec:modelmanipulation}
The model training stage involves creating an effective model replicating the expected relationship between the inputs in the training dataset and the model’s outputs. Thus, manipulating the model directly to remove specific relationships may be a good way to unlearn samples. In this section, we comprehensively review the state-of-the-art studies on unlearning through model manipulation. Again, the verification techniques are discussed separately for each category.

\subsection{Manipulation Based on Model Shifting}
\label{subsec:modelshifting}
\subsubsection{Unlearning Schemes Based on Model Shifting}~

\begin{figure}
  \centering
  \includegraphics[width=0.7\linewidth]{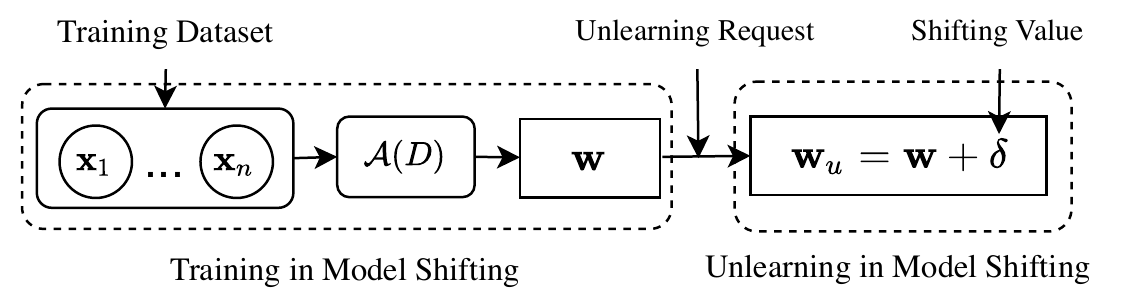}
  \caption{Unlearning Schemes Based on Model Shifting.}
  \label{fig:modelshifting}
\end{figure}

As shown in Figure~\ref{fig:modelshifting}, model shifting methods usually eliminate the influence of unlearning data by directly updating the model parameters. These methods mainly fall into one of two types – influence unlearning and Fisher unlearning – but there are a few other methods.

\emph{(1). Influence unlearning methods}

Influence unlearning methods are usually based on influence theory~\cite{DBLP:conf/icml/KohL17}. Guo et al.~\cite{DBLP:conf/icml/GuoGHM20} proposed a novel unlearning scheme called \textit{certified removal}. Inspired by differential privacy~\cite{DBLP:journals/jmlr/0015GS020}, \textit{certified removal} first limits the maximum difference between the unlearned and retrained models. Then, by applying a single step of Newton’s method on the model parameters, a \textit{certified removal} mechanism is provided for practical applications of  $L_{2}-$ regularized linear models that are trained using a differentiable convex loss function. Additionally, the training loss is perturbed with a loss perturbation technique that hides the \textit{gradient residual}. This further prevents any adversaries from extracting information from the unlearned model. It is worth noting, however, that this solution is only applicable to simple machine learning models, such as linear models, or only adjusts the linear decision-making layer for deep neural networks, which does not eliminate the information of the removed data sample since the representations are still learned within the model.

Izzo et al.~\cite{DBLP:conf/aistats/IzzoSCZ21} proposed an unlearning method based on a gradient update called \textit{projection residual update} (PRU). The method focuses on linear regression and shows how to improve the algorithm's run-time given in~\cite{DBLP:conf/icml/GuoGHM20} from quadratic complexity to linear complexity. The unlearning intuition is as follows: if one can calculate the values $\hat{y}_{i_{\mathcal{D}_u}}=w_{u} \left(x_{i_{\mathcal{D}_u}}\right)$, predicted by the unlearned model on each of the unlearned samples $x_{i_{\mathcal{D}_u}}$ in $\mathcal{D}_u$ without knowing $w_{u}$, and then minimize the loss of already-trained model on the synthetic samples $\left(x_{i_{\mathcal{D}_u}}, \hat{y}_{i}\right)$, the parameters will move closer to $\mathbf{w}_{u}$ since it will achieve the minimum loss with samples $\left(x_{i_{\mathcal{D}_u}}, \hat{y}_{i_{\mathcal{D}_u}}\right)$. To calculate the values $\hat{y}_{i_{\mathcal{D}_u}}$ without knowing $w_{u}$, they introduced a statistics technique and computed leave-one-out residuals. Similar to the above, this method only considers the unlearning process in simple models.

Information leaks may not only manifest in a single data sample but also in groups of features and labels ~\cite{DBLP:journals/corr/abs-2108-11577}. For example, a user’s private data, such as their telephone number and place of residence, are collected by data providers multiple times and generated as different samples of the training dataset. Therefore, unlearning operations should also focus on unlearning a group of features and corresponding labels.

To solve such problems, Warnecke et al.~\cite{DBLP:journals/corr/abs-2108-11577} proposed a \textit{certified unlearning} scheme for unlearning features and labels. By reformulating the influence estimation of samples on the already-trained models as a form of unlearning, they derived a versatile approach that maps changes of the training dataset in retrospection to closed-form updates of the model parameters. They then proposed different unlearning methods based on \textit{first-order} and \textit{second-order} gradient updates for two different types of machine learning models. For the \textit{first-order} update, the parameters were updated based on the difference between the gradient of the original and the perturbed samples. For the \textit{second-order} update, they approximated an inverse Hessian matrix based on the scheme proposed in~\cite{DBLP:journals/jmlr/AgarwalBH17} and updated the model parameters based on this approximate matrix. Theoretical guarantees were also provided for feature and label unlearning by extending the concept of differential privacy~\cite{DBLP:journals/jmlr/0015GS020} and certified unlearning~\cite{DBLP:conf/icml/GuoGHM20}. However, this solution is only suitable for feature unlearning from tabular data and does not provide any effective solution for image features.

\emph{(2). Fisher unlearning method}

The second type of model shifting technique uses the Fisher information~\cite{DBLP:journals/jmlr/Martens20} of the remaining dataset to unlearn specific samples, with noise injected to optimize the shifting effect. Golatkar et al.~\cite{DBLP:conf/cvpr/GolatkarAS20} proposed a weight \textit{scrubbing} method to unlearn information about a particular class as a whole or a subset of samples within a class. They first give a computable upper bound to the amount of the information retained about the unlearning dataset after applying the unlearning procedure, which is based on the Kullback-Leibler (KL) divergence and Shannon mutual information. Then, an optimal quadratic unlearning algorithm based on a Newton update and a more robust unlearning procedure based on a noisy Newton update were proposed. Both schemes can ensure that a cohort can be unlearned while maintaining good accuracy for the remaining samples. However, this unlearning scheme is based on various assumptions, which limits its applicability.

For deep learning models, bounding the information that can be extracted from the perspective of weight or weight distribution is usually complex and may be too restrictive. Deep networks have a large number of equivalent solutions in the distribution space, which will provide the same activation on all test samples~\cite{DBLP:conf/eccv/GolatkarAS20}. Therefore, many schemes have redirected unlearning operations from focusing on the weights to focus on the final activation. 

Unlike their previous work, Golatkar et al.~\cite{DBLP:conf/eccv/GolatkarAS20} provide bounds for how much information can be extracted from the final activation. They first transformed the bounding from a weight perspective to final activation based on Shannon mutual information and proposed a computable bound using the $KL$-divergence between the distribution of final activation of an unlearned model and retrained model. Inspired by the neural tangent kernel (NTK)~\cite{DBLP:conf/stoc/JacotGH21,DBLP:conf/icml/HuangL0021}, they considered that deep network activations can be approximated as a linear function of the weights. Hence, an optimal unlearning procedure is then provided based on a Fisher information matrix. However, due to the specific structure of deep neural networks, considering unlearning process only in the final activation layer may not satisfy the effectiveness of unlearning. Once an attacker obtains all model parameters in a white-box scenario, they can still infer information from the middle layers.

Golatkar et al.~\cite{DBLP:conf/cvpr/GolatkarARPS21} also proposed a mix-privacy unlearning scheme based on a new \textit{mixed-privacy} training process. This new training process assumes the traditional training dataset can be divided into two parts: \textit{core} data and \textit{user} data. Model training on the \textit{core} data is non-convex, and then further training, based on the quadratic loss function, is done with the \textit{user} data to meet the needs of specific user tasks. Based on this assumption, unlearning operations on the \textit{user} data can be well executed based on the existing quadratic unlearning schemes. Finally, they also derived bounds on the amount of information that an attacker can extract from the model weights based on mutual information. Nevertheless, the assumption that the training dataset is divided into two parts and that the model is trained using different methods on each of these parts restricts unlearning requests to only those data that are easy to unlearn, making it difficult to unlearn other parts of the data.

Liu et al.~\cite{DBLP:conf/infocom/LiuXYWL22} transferred the unlearning method from a centralized environment to federated learning by proposing a distributed Newton-type model updating algorithm to approximate the loss function trained by the local optimizer on the remaining dataset. This method is based on the Quasi-Newton method and uses a first-order Taylor expansion. They also use diagonal empirical Fisher Information Matrix (FIM) to efficiently and accurately approximate the inverse Hessian vector, rather than computing it directly, to further reduce the cost of the retraining process. However, this solution will result in a significant reduction in accuracy when dealing with complex models.

\emph{(3). Other Shifting Schemes}

Schelter et al.~\cite{DBLP:conf/cidr/Schelter20} introduced the problem of making trained machine learning models unlearn data via \textit{decremental updates}. They described three decremental update algorithms for different machine learning tasks. These included one based on item-based collaborative filtering, another based on ridge regression, and the last based on $k$-nearest neighbors. With each machine learning algorithm, the intermediate results are retained, and the model parameters are updated based on the intermediate results and unlearning data $D_{u}$, resulting in an unlearned model. However, this strategy can only be utilized with those models that can be straightforwardly computed to obtain the model parameters after the unlearning process, limiting the applicability of this scheme.

In addition, Graves et al.~\cite{DBLP:conf/aaai/GravesNG21} also proposed a laser-focused removal of sensitive data, called \textit{amnesiac unlearning}. During training, the model provider retains a variable that stores which samples appear in which batch, as well as the parameter updates for each batch. When a data unlearning request arrives, the model owner undoes the parameter updates from only the batches containing the sensitive data, that is  $\mathcal{M}_{w_u}=\mathcal{M}_{w}-\sum \Delta_{w}$, where $\mathcal{M}_{w}$ is the already-trained model and $\Delta_{w}$ are the parameter updates after each batch. Because undoing some parameters might greatly reduce the performance of the model, the model provider can perform a small amount of fine-tuning after an unlearning operation to regain performance. This approach requires the storage of a substantial amount of intermediate data. As the storage interval decreases, the amount of cached data increases, and smaller intervals lead to more efficient model unlearning. Therefore, a trade-off exists between efficiency and effectiveness in this method.

The above methods mainly focused on the core problem of empirical risk minimization, where the goal is to find approximate minimizers of the empirical loss on the remaining training dataset after unlearning samples ~\cite{DBLP:conf/icml/GuoGHM20,DBLP:conf/aistats/IzzoSCZ21}. Sekhari et al.~\cite{DBLP:journals/corr/abs-2103-03279} proposed a more general method of reducing the loss of unseen samples after an unlearning process. They produced an unlearned model by removing the contribution of some samples from an already-trained model using a disturbance update calculated based on some cheap-to-store data statistics during training. In addition, they proposed an evaluation parameter to measure the unlearning capacity. They also improved the data unlearning capacity of convex loss functions, which saw a quadratic improvement in terms of the dependence of \textit{d} over differential privacy, where \textit{d} is the problem dimension.

\subsubsection{Verifiability of Schemes Based on Parameter Shifting}~

Izzo et al.~\cite{DBLP:conf/aistats/IzzoSCZ21} provided two metrics to measure the effectiveness: $L_2$ \textit{distance} and \textit{feature injection test}. $L_2$ \textit{distance} measures the distance between the unlearned model and the retrained model. If the $L_2$ \textit{distance} is small, the models are guaranteed to make similar predictions, which could reduce the impact of output-based attacks, like a membership inference attack. The \textit{feature injection test} can be thought of as a verification scheme based on a poisoning attack.

Golatkar et al.~\cite{DBLP:conf/cvpr/GolatkarAS20, DBLP:conf/eccv/GolatkarAS20,DBLP:conf/cvpr/GolatkarARPS21} verify the effectiveness of their unlearning schemes based on accuracy and relearning time. They also developed two new verification metrics: \textit{model confidence} and \textit{information bound}~\cite{DBLP:conf/cvpr/GolatkarAS20}. \textit{Model confidence} is formulated by measuring the distribution of the entropy of the output predictions on the remaining dataset, the unlearning dataset, and the test dataset. Then they evaluated the similarity of those distributions against the confidence of a trained model that has never seen the unlearning dataset. The higher the degree of similarity, the better the effect of the unlearning process. The \textit{information bound} metric relies on KL-divergence to measure the information remaining about the unlearning dataset within the model after the unlearning process.

Different from their previous work, Golatkar et al.~\cite{DBLP:conf/eccv/GolatkarAS20} also evaluate the information remaining within the weights and the activation. In their other work~\cite{DBLP:conf/cvpr/GolatkarARPS21}, they provided a new metric, \textit{activation distance}, to analyze the distance between the final activations of an unlearned model and a retrained model. This is a similar metric to \textit{model confidence} ~\cite{DBLP:conf/cvpr/GolatkarAS20}. In addition, they also use attack-based methods for verification ~\cite{DBLP:conf/eccv/GolatkarAS20,DBLP:conf/cvpr/GolatkarARPS21}. 

Guo et al.~\cite{DBLP:conf/icml/GuoGHM20}, Warnecke et al.~\cite{DBLP:journals/corr/abs-2108-11577} and Sekhari et al.~\cite{DBLP:journals/corr/abs-2103-03279} provide a method of theoretical verification to verify the effectiveness of their proposed unlearning schemes. Based on the guarantee provided by \textit{certified unlearning}, they limit the distribution similarity between the unlearned model and the retrained model.  Warnecke et al.~\cite{DBLP:journals/corr/abs-2108-11577} also use the \textit{exposure metric}~\cite{DBLP:conf/uss/Carlini0EKS19} to measure the remaining information after unlearning. Liu et al.~\cite{DBLP:conf/infocom/LiuXYWL22} analyzed the validity of the unlearning scheme through two aspects. The first metric, Symmetric Absolute Percentage Error (SAPE), is created based on accuracy. The second metric is the difference between the distribution of the model after the unlearning process and the distribution of the retraining model.

\subsection{Manipulation Based on Model Pruning}
\label{subsec:modelpruning}
\subsubsection{Unlearning Schemes Based on Model Pruning}~

\begin{figure}
  \centering
  \includegraphics[width=0.9\linewidth]{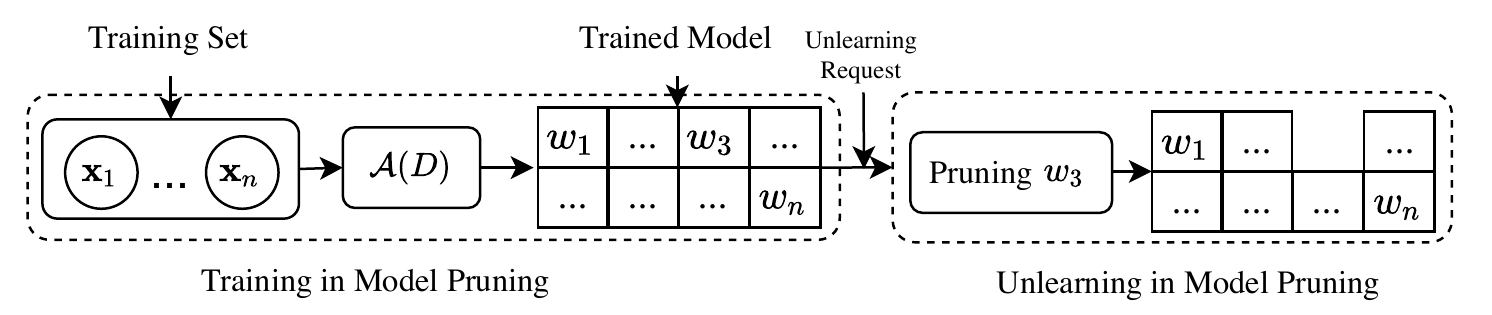}
  \caption{Unlearning Schemes Based on Model Pruning.}
  \label{fig:modelpruning}
\end{figure}

As shown in Figure~\ref{fig:modelpruning}, methods based on model pruning usually prune a trained model to produce a model that can meet the requests of unlearning. It is usually applied in the scenario of federated learning, where a model provider can modify the model’s historical parameters as an update. Federated learning is a distributed machine learning framework that can train a unified deep learning model across multiple decentralized nodes, where each node holds its own local data samples for training, and those samples never need to be exchanged with any other nodes~\cite{DBLP:journals/comsur/LimLHJLYNM20}. There are mainly three types of federated learning: \textit{horizontal}, \textit{vertical}, and \textit{transfer learning}~\cite{DBLP:journals/iotj/ImteajTWLA22}.

Based on the idea of trading the central server’s storage for the unlearned model’s construction, Liu et al.~\cite{DBLP:conf/iwqos/LiuMYWL21} proposed an efficient federated unlearning methodology, \textit{FedEraser}. Historical parameter updates from the clients are stored in the central server during the training process, and then the unlearning process unfolds in four steps: (1) \textit{calibration training}, (2) \textit{update calibrating}, (3) \textit{calibrated update aggregating}, and (4) \textit{unlearned model updating}, to achieve the unlearning purpose.  In \textit{calibration training} and \textit{update calibration} steps, several rounds of a calibration retraining process are performed to approximate the unlearning updates without the target client. In the \textit{calibrated update aggregating} and the \textit{unlearned model updating} steps, standard federated learning aggregation operations are used to aggregate those unlearning updates and further update the global model. This eliminates the influence of the target data. 

However, the effectiveness of this scheme will decrease dramatically as the number of unlearning requests increases; this is because the gradients are cached during the training phase, and the unlearning process will not update these gradients to satisfy subsequent unlearning requests~\cite{DBLP:conf/iwqos/LiuMYWL21}. Second, this solution also requires caching of intermediate data, which will cost more storage.

Inspired by the observation that different channels have a varying contribution to different classes in trained CNN models. Wang et al.~\cite{DBLP:conf/www/Wang0XQ22} analyzed the problem of selectively unlearning classes in a federated learning setting. They introduced the concept of term frequency-inverse document frequency (TF-IDF)~\cite{DBLP:journals/asc/ThakkarC20} to quantify the class discrimination of the channels. Similar to analyzing how relevant a word is to a document in a set of documents, they regarded the output of a channel as a word and the feature map of a category as a document. Channels with high TF-IDF scores have more discriminatory power in the target categories and thus need to be pruned. An unlearning procedure via channel pruning~\cite{DBLP:conf/cvpr/GuoWLY20} was also provided, followed by a fine-tuning process to recover the performance of the pruned model. In their unlearning scheme, however, while the parameters associated with the class that needs to be unlearned are pruned, the parameters with other classes also become incomplete, which will affect the model performance. Therefore, the unlearned model is only available when the fined-tuned training process is complete.

Baumhauer et al.~\cite{DBLP:journals/corr/abs-2002-02730} provided a machine unlearning scheme based on linear filtration. They first transformed the existing logit-based classifier models into an integrated model which can be decomposed into a (potentially nonlinear) feature extraction, followed by a multinomial logistic regression. Then, they focused the unlearning operation on the logistic regression layer, proposing a “black-box” unlearning definition. To unlearn the given samples, four different filtration methods are defined, namely, \textit{naive unlearning}, \textit{normalization}, \textit{randomization}, and \textit{zeroing}. These effectively filter the outputs of the logistic regression layer. On the contrary, they only considered the unlearning process within the last layer, which will lead to a potential risk that if an attacker gets access to the model parameters of the middle layer, the information of unlearning data may also be leaked.

\subsubsection{Verifiability of Schemes based on Model Pruning}~

Liu et al.~\cite{DBLP:conf/iwqos/LiuMYWL21} present an experimental verification method based on a membership inference attack. Two evaluation parameters are specified: \textit{attack precision} and \textit{attack recall}, where \textit{attack precision} denotes the proportion of unlearned samples that is expected to participate in the training process. \textit{Attack recall} denotes the fraction of unlearned samples that can be correctly inferred as part of the training dataset. In addition, a \textit{prediction difference} metric is also provided, which measures the difference in prediction probabilities between the original global model and the unlearned model.  Wang et al.~\cite{DBLP:conf/www/Wang0XQ22} evaluate verifiability based on model accuracy.

Baumhauer et al.~\cite{DBLP:journals/corr/abs-2002-02730} defined a divergence measure based on a Bayes error rate for evaluating the similarity of the resulting distributions $P\left(\mathbf{L}_{\text {seen }}\right)$ and $P\left(\mathbf{L}_{\neg \text { seen }}\right)$, where $\mathbf{L}_{\text {seen }}$ and $\mathbf{L}_{\neg \text { seen }}$  are the pre-softmax outputs of the unlearned model and a retrained model. When the result of the Bayes error rate is close to $0$, it indicates that $P\left(\mathbf{L}_{\text {seen }}\right)$ and $P\left(\mathbf{L}_{\text {-seen }}\right)$ are similar and the unlearning process has unlearned the sample’s information from the model. In addition, they also use a model inversion attack to evaluate verifiability ~\cite{DBLP:conf/ccs/FredriksonJR15}.

\subsection{Manipulation Based on Model Replacement}
\label{subsec:modelreplacement}
\subsubsection{Unlearning Schemes Based on Model Replacement}~

\begin{figure}
  \centering
  \includegraphics[width=.85\linewidth]{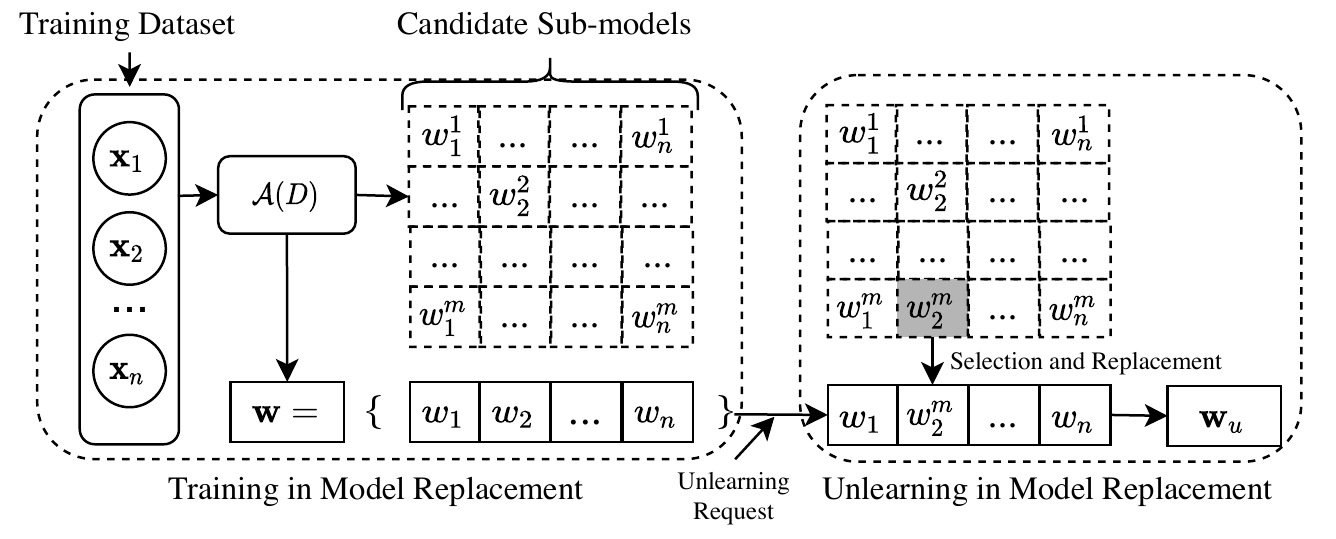}
  \caption{Unlearning Schemes Based on Model Replacement.}
  \label{fig:modelreplacement}
\end{figure}

As shown in Figure~\ref{fig:modelreplacement}, model replacement-based methods usually calculate almost all possible sub-models in advance during the training process and store them together with the deployed model. Then, when an unlearning request arrives, only the sub-models affected by the unlearning operation need to be replaced with the pre-stored sub-models. This type of solution is usually suitable for some machine learning models, such as tree-based models. Decision tree is a tree-based learning model, in which each leaf node represents a prediction value, and each internal node is a decision node associated with an attribute and threshold value. Random forest is an integrated decision tree model that aims to improve prediction performance~\cite{DBLP:journals/pami/ShenGWZWY21,DBLP:journals/ai/MaliahS21}. 

To improve the efficiency of the unlearning process for tree-based machine learning models, Schelter et al.~\cite{DBLP:conf/sigmod/SchelterGD21} proposed \textit{Hedgecut}, a classification model based on extremely randomized trees (ERTs)~\cite{DBLP:journals/isci/ChengSWGX21}. First, during the training process, the tree model is divided into robust splits and non-robust splits based on the proposed robustness quantification factor. A robust split indicates that the subtree’s structure will not change after unlearning a small number of samples, while for non-robust splits, the structure may be changed. In the case of unlearning a training sample, \textit{HedgeCut} will not revise robust splits, but will update those leaf statistics. For non-robust splits, \textit{HedgeCut} recomputes the split criterion of the maintained subtree variants, which were previously kept inactive, and selects a subtree variant as a new non-robust split of the current model. 

For the tree-based models, Brophy et al.~\cite{DBLP:conf/icml/BrophyL21} also proposed DaRE (Data Removal-Enabled) forests, a random forest variant that enables the efficient removal of training samples. DaRE is mainly based on the idea of retraining subtrees only as needed. Before the unlearning process, most $k$ randomly-selected thresholds per attribute are computed, and intermediate statistics data are stored within each node in advance. This information is sufficient to recompute the split criterion of each threshold without iterating through the data, which can greatly reduce the cost of recalculation when unlearning the dataset. They also introduced random nodes at the top of each tree. Intuitively, the nodes near the top of the tree affect more samples than those near the bottom, which makes it more expensive to retrain them when necessary. Random nodes minimally depend on the statistics of the data, rather than the way greedy methods are used, and rarely need to be retrained. Therefore, random nodes can further improve the efficiency of unlearning. 

The above two schemes need to compute a large number of possible tree structures in advance, which would cost a large number of storage resources~\cite{DBLP:conf/sigmod/SchelterGD21, DBLP:conf/icml/BrophyL21}. Besides, this replacement scheme is difficult to be applied to other machine learning models, such as deep learning models, since it is difficult to achieve partial model structure after removing each sample in advance.

Chen et al.~\cite{DBLP:journals/corr/abs-2111-11869} proposed a machine unlearning scheme called \textit{WGAN} unlearning, which removes information by reducing the output confidence of unlearned samples. Machine learning models usually have different confidence levels toward the model’s outputs~\cite{DBLP:conf/uss/CarliniTWJHLRBS21}. To reduce confidence, \textit{WGAN} unlearning first initializes a generator as the trained model that needs to unlearn data. Then, the generator and discriminator are trained alternatingly until the discriminator cannot distinguish the output difference of the model between unlearning dataset and third-party data. Until this, the generator then becomes the final unlearned model. However, this method achieves unlearning process through an alternating training process, which brings a limited improvement in efficiency compared to the unlearning method of retraining from scratch.

Wu et al.~\cite{DBLP:conf/icml/WuDD20} proposed an approximate unlearning method based on intermediate parameters cached during the training phase called \textit{DeltaGrad}, which could quickly unlearn information from machine learning models that are based on gradient descent algorithms. They divided the retraining process into two parts. One part computes the full gradients exactly based on the remaining training dataset. The other part uses the L-BGFS algorithm~\cite{DBLP:journals/jmlr/MokhtariR15} and a set of updates from some prior iterations to calculate Quasi-Hessians approximating the true Hessian-vector. These Quasi-Hessians are then used to approximate the update in the remaining process. These two parts train cooperatively to generate the unlearned model. This approach will reduce the performance of the model, however, after unlearning process since part of the model update is calculated based on the approximative methods. In addition, the number of iterations required for the model to converge will also increase, which will reduce the efficiency of the unlearning process.

\subsubsection{Verifiability of Schemes Based on Model Replacement}~

Chen et al.~\cite{DBLP:journals/corr/abs-2111-11869} verified their proposed scheme with a membership inference attack and a technique based on false negative rates (FNRs)~\cite{DBLP:conf/icassp/KoizumiMHSU19}, where $FNR$: $FNR=\frac{FN}{TP+ FN}$, $TP$ means that the membership inference attack test samples were considered to be training dataset and $FN$ means the data was deemed to be non-training data. If the target model successfully unlearns the samples, the member inference attack will treat the training dataset as non-training data. Thus $FN$ will be large, while $TP$ will be small, and the corresponding $FNR$ will be large. Indirectly, this reflects the effectiveness of the unlearning process.

Schelter et al.~\cite{DBLP:conf/sigmod/SchelterGD21}, Brophy et al.~\cite{DBLP:conf/icml/BrophyL21} and Wu et al.~\cite{DBLP:conf/icml/WuDD20} only provide evaluations in terms of runtime and accuracy, and they do not provide reasonable experimental or theoretical verifiability guarantees of their unlearning processes.

\setlength\rotFPtop{0pt plus 1fil} 
\begin{sidewaystable}
    \caption{The Surveyed Studies that Employed Model Manipulation Techniques for Unlearning Process.}
    \label{tab:manipulation}
	\resizebox{\textwidth}{!}{%
	\begin{tabular}{cccccccccc}
	\toprule
	Papers 
	& \begin{tabular}{c}Unlearning \\Methods\end{tabular}
	& \begin{tabular}{c}Unlearning \\Target  \end{tabular}
	& Training Dataset 
	& Intermediates 
	&  \begin{tabular}{c}Unlearned Samples' \\Type \end{tabular}
	& \begin{tabular}{c}Target Models'  \\Type   \end{tabular}
	& Consistency 
	& Accuracy 
	& Verifiability \\  
	\midrule

	\rowcolor{gray!40}
	Guo et al.~\cite{DBLP:conf/icml/GuoGHM20}   
	& \begin{tabular}{c}Model \\Shifting\end{tabular}
	& \begin{tabular}{c}Strong\\ Unlearning  \end{tabular}          
	& Yes 
	& No  
	& Samples                                              
	& \begin{tabular}{c}Linear Models with \\Strongly Convex  \\ Regularization  \end{tabular} 
	& No  
	& No  
	& \begin{tabular}{c}Theory-Based \end{tabular} \\                                                                    
	
	Izzo et al.~\cite{DBLP:conf/aistats/IzzoSCZ21}
	& \begin{tabular}{c}Model \\Shifting \end{tabular}
	& \begin{tabular}{c}Strong\\ Unlearning \end{tabular}
	& No  
	& Yes 
	& Batches                                              
	& \begin{tabular}{c}Linear and \\Logistic Regression  Models\end{tabular}               
	& No  
	& No  
	& \begin{tabular}{c}$L^2$ distance and \\Attack-Based \end{tabular}      \\                                                         
	
	\rowcolor{gray!40}
	Warnecke et al.~\cite{DBLP:journals/corr/abs-2108-11577}  
	& \begin{tabular}{c}Model \\Shifting \end{tabular}
	& \begin{tabular}{c}Strong\\ Unlearning  \end{tabular}
	& Yes 
	& No  
	& \begin{tabular}{c}Features\\and Labels \end{tabular} 
	& \begin{tabular}{c}Convex  or \\Non-convex models \end{tabular}                 
	& No  
	& No  
	& \begin{tabular}{c} Theory-Based and Method in ~\cite{DBLP:conf/uss/Carlini0EKS19}  \end{tabular}       \\          
	
	Golatkar et al.~\cite{DBLP:conf/cvpr/GolatkarAS20}        
	& \begin{tabular}{c}Model \\Shifting \end{tabular}
	&\begin{tabular}{c} Strong\\ Unlearning\end{tabular}
	& Yes 
	& No  
	&\begin{tabular}{c}Samples \\in One Class \end{tabular}                                   
	& DNN                                                                                    
	& No  
	& No  
	& \begin{tabular}{c}Accuracy-based, \\ Relearn time-based, \\ Model confidence and \\Information Bound-Based \end{tabular} \\
	
	\rowcolor{gray!40}
	Golatkar et al.~\cite{DBLP:conf/eccv/GolatkarAS20} 
	& \begin{tabular}{c}Model \\Shifting    \end{tabular}
    & \begin{tabular}{c}Strong \\Unlearning     \end{tabular} 
	& Yes 
	& No  
	& Samples                                              
	& DNN                                                                                    
	& No  
	& No  
	&\begin{tabular}{c}Accuracy-based,\\ Relearn time-based,\\ Attack-based  and \\ Information Bound-Based\end{tabular}      \\   

	Liu et al.~\cite{DBLP:conf/infocom/LiuXYWL22}
	& \begin{tabular}{c}Model \\Shifting    \end{tabular}
        & \begin{tabular}{c}Strong \\Unlearning     \end{tabular} 
	& Yes 
	& Yes  
	& Samples                                              
	& DNN                                                                                    
	& No  
	& No  
	&\begin{tabular}{c}Accuracy-based\end{tabular}      \\   
        
 	\rowcolor{gray!40}
	Golatkar et al.~\cite{DBLP:conf/cvpr/GolatkarARPS21}   
	& \begin{tabular}{c}Model \\Shifting  \end{tabular}
	& \begin{tabular}{c}Strong \\Unlearning  \end{tabular}
	& Yes 
	& No  
	& Samples                                              
	& DNN                                                                                    
	& No  
	& No  
	&\begin{tabular}{c}Accuracy-based, \\Relearn time-based,\\ Activation distance and \\ Attack-based \end{tabular}      \\

	Schelter et al.~\cite{DBLP:conf/cidr/Schelter20}   
	& \begin{tabular}{c}Model \\Shifting \end{tabular}
	& \begin{tabular}{c}Exact \\Unlearning    \end{tabular}    
	& No  
	& Yes 
	& Samples                                              
	& Specified model                                                                        
	& Yes 
	& Yes 
	& - \\

 	\rowcolor{gray!40}
	Graves et al.~\cite{DBLP:conf/aaai/GravesNG21}  
	& \begin{tabular}{c}Model \\Shifting   \end{tabular}
	& \begin{tabular}{c}Strong \\Unlearning    \end{tabular}     
	& No  
	& Yes 
	& Samples                                              
	& DNN                                                                                    
	& No  
	& No  
	& Attack-Based \\
 	
	Sekhari et al.~\cite{DBLP:journals/corr/abs-2103-03279}
	& \begin{tabular}{c}Model \\Shifting   \end{tabular}
	& \begin{tabular}{c}Strong \\Unlearning \end{tabular}
	& No  
	& Yes 
	& Samples                                              
	& Convex Models                                                                          
	& No  
	& No 
	& \begin{tabular}{c}Theory-Based\end{tabular}       \\                                                                          
	
	\rowcolor{gray!40}
	Wang et al.~\cite{DBLP:conf/www/Wang0XQ22}    
	& \begin{tabular}{c}Model \\Pruning  \end{tabular}
	&\begin{tabular}{c} Strong \\Unlearning   \end{tabular}
	& Yes 
	& No  
	& Client Data                                          
	& Federated Learning Model                                                               
	& No  
	& No  
	& -         \\                                                                      
	
	Baumhauer et al.~\cite{DBLP:journals/corr/abs-2002-02730} 
	& \begin{tabular}{c}Model \\Pruning \end{tabular}
	& \begin{tabular}{c}Weak \\Unlearning \end{tabular}
	& No  
	& No  
	& Classes                                              
	& Logit-Based Classifiers                                                                
	& No  
	& No  
	& Attack-Based  \\
	
    \rowcolor{gray!40}	
	Schelter et al~\cite{DBLP:conf/sigmod/SchelterGD21}     
    & \begin{tabular}{c}Model \\Replacement  \end{tabular}
 	& \begin{tabular}{c}Exact \\Unlearning \end{tabular}
	& No  
	& Yes 
	& Samples                                              
	& Extremely Randomized Trees                                                             
	& Yes 
	& Yes
	& -  \\
	
	Brophy et al~\cite{DBLP:conf/icml/BrophyL21}    
	& \begin{tabular}{c}Model \\Replacement   \end{tabular}
	& \begin{tabular}{c}Exact \\Unlearning      \end{tabular}   
	& Yes 
	& Yes 
	& Batches                                              
	& Random Forests                                                                         
	& Yes 
	& Yes 
	& - \\
  
    \rowcolor{gray!40}		
	Chen et al.~\cite{DBLP:journals/corr/abs-2111-11869}  
    & \begin{tabular}{c}Model \\Replacement     \end{tabular}
 	& \begin{tabular}{c}Strong \\Unlearning  \end{tabular}
	& No  
	& No  
	& Samples                                              
	& Deep Classifier Models                                                                 
	& No  
	& No  
	& Attack-Based   \\

	Wu et al.~\cite{DBLP:conf/icml/WuDD20} 
    & \begin{tabular}{c}Model \\Replacement \end{tabular}
    & \begin{tabular}{c}Strong \\Unlearning  \end{tabular}
	& Yes 
	& Yes
	& Samples                                              
	& SGD-Based Models                                                                       
	& No  
	& No  
	& -      \\                                                                           

	\bottomrule
	\end{tabular}%
	}
\end{sidewaystable}

\subsection{Summary of Model Manipulation}
 
In these last subsections, we reviewed studies that apply model shifting, model pruning, and model replacement techniques as unlearning processes. A summary of the surveyed studies is shown in Table~\ref{tab:manipulation}, where we list the key differences between each paper.

Compared to the unlearning schemes based on data reorganization, we can see that few of the above papers make use of intermediate data for unlearning. This is because the basic idea of those unlearning schemes is to directly manipulate the model itself, rather than the training dataset. The model manipulation methods calculate the influence of each sample and offset that influence using a range of techniques~\cite{DBLP:conf/icml/KohL17}, while data reorganization schemes usually reorganize the training dataset to simplify the unlearning process. For this reason, model manipulation methods somewhat reduce the resource consumption used by intermediate storage.

Second, most of the above schemes focus on relatively simple machine learning problems, such as linear logistic regression, or complex models with special assumptions~\cite{DBLP:conf/icml/GuoGHM20, DBLP:conf/aistats/IzzoSCZ21,DBLP:conf/cvpr/GolatkarAS20, DBLP:conf/eccv/GolatkarAS20}. Removing information from the weights of standard convolutional networks is still an open problem, and some preliminary results are only applicable to small-scale problems. One of the main challenges with unlearning processes for deep networks is how to estimate the impact of a given training sample on the model parameters. Also, the highly non-convex losses of CNNs make it very difficult to analyze those impacts on the optimization trajectory. Current research has focused on simpler convex learning problems, such as linear or logistic regression, for which theoretical analysis is feasible. Therefore, evaluating the impact of specific samples on deep learning models and further proposing unlearning schemes for those models are two urgent research problems.

In addition, most model manipulation-based methods will affect the consistency or prediction accuracy of the original models. There are two main reasons for this problem. First, due to the complexity of calculating the impact of the specified sample on the model, manipulating a model’s parameters based on unreliable impact results or assumptions will lead to a decline in model accuracy. Secondly, Wang et al.’s~\cite{DBLP:conf/www/Wang0XQ22} scheme pruned specific parameters in the original models, which will also reduce the accuracy of the model due to the lack of some model prediction information. Thus, more efficient unlearning mechanisms, which simultaneously ensure the validity of the unlearning process and guarantee performance, are worthy of research.

It is worth pointing out that most schemes provide a reasonable method with which to evaluate the effectiveness of the unlearning process. Significantly, model manipulation methods usually give a verifiability guarantee using theory-based and information bound-based methods~\cite{DBLP:conf/icml/GuoGHM20, DBLP:conf/cvpr/GolatkarAS20, DBLP:conf/eccv/GolatkarAS20}. Compared to the simple verification methods based on accuracy, relearning, or attacks, the methods based on theory or information bounds are more effective. This is because simple verification methods usually verify effectiveness based on output confidence. While the effects of the 
samples to be unlearned can be hidden from the output of the network, insights may still be gleaned by probing deep into its weights. Therefore, calculating and limiting the maximum amount of information that may be leaked at the theoretical level will be a more convincing method. Overall, however, more theory-based techniques for evaluating verifiability are needed.

In summary, the unlearning methods based on model shifting usually aim to offer higher efficiency by making certain assumptions about the training process, such as which training dataset or optimization techniques have been used. In addition, those mechanisms that are effective for simple models, such as linear regression models, become more complex when faced with advanced deep neural networks. Model pruning schemes require far-reaching modifications of the existing architecture of the model in the unlearning process ~\cite{DBLP:conf/www/Wang0XQ22, DBLP:journals/corr/abs-2002-02730}, which could affect the performance of the unlearned models. It is worth noting that model replacement unlearning methods usually need to calculate all possible parameters and store them in advance, since they unlearn by quickly replacing the model parameters using these pre-calculated parameters. Thus, more effective unlearning schemes, that simultaneously consider model usability, storage costs, and the applicability of the unlearning process, are urgent research problems. 

\section{Open questions and future directions}

\label{sec:outlookandfuture}
In this section, we will analyze current and potential trends in machine unlearning, and summarize our findings. In addition, we identify several unanswered research directions that could be addressed to progress the foundation of machine unlearning and shape the future of AI.

\subsection{Open Questions}

As research continues to evolve, machine unlearning may expand further in the following areas, and this potential trend has already begun to take shape.

\subsubsection{The universality of unlearning solutions}~

Unlearning schemes with higher compatibility need to be explored. As development progresses, machine unlearning schemes supporting different models and unlearning data types have been proposed in various fields. For example, Zhang et al.~\cite{DBLP:conf/mm/ZhangBHX22} provided an unlearning scheme in image retrieval, while Chen et al.~\cite{DBLP:conf/ccs/Chen000H022} considered graph unlearning problem. However, most of the current unlearning schemes are limited to a specific scenario. They are mostly designed to leverage the special characteristics of a particular learning process or training scheme~\cite{DBLP:conf/ccs/Chen000H022,DBLP:conf/cidr/Schelter20,DBLP:conf/iwqos/LiuMYWL21}.  Although it is feasible to design an appropriate unlearning scheme for every model, this is an inefficient approach that would require many manual interventions~\cite{DBLP:conf/aaai/WuHS22,DBLP:journals/corr/abs-2201-05629}.

Therefore, universality unlearning schemes should be not only applicable to different model structures and training methods, but also to different types of training datasets, such as graphs, images, text, or audio data. The data pruning-based scheme is an existing and effective approach that could achieve universality unlearning purposes based on ensemble learning techniques~\cite{DBLP:conf/sp/BourtouleCCJTZL21}. However, this method breaks the correlation relationships in some scenarios, which is not suitable for models that require correlation information to complete training.
    
\subsubsection{The security of machine unlearning}~

Unlearning schemes should ensure the security of any data, especially the unlearned dataset.  Recently, existing research has shown that the unlearning operation not only does not reduce the risk of user privacy leakage but actually increases this risk~\cite{DBLP:conf/ccs/BeguelinWTRPOKB20, DBLP:journals/corr/abs-1912-07942}. These attack schemes mainly compare the models before and after the unlearning process. Thus, a membership inference attack or a poisoning attack would reveal a significant amount of detailed information about the unlearned samples~\cite{DBLP:conf/ccs/Chen000HZ21,DBLP:conf/aaai/MarchantRA22}. In order to counteract such attacks, Neel et al.~\cite{DBLP:conf/alt/Neel0S21} have proposed a protection method based on Gaussian perturbation in their unlearning scheme.

In addition, many previous unlearning schemes rely on the remaining dataset, intermediate cached model’s parameters. However, they do not consider the security of this intermediate information and whether an attack would recover any information about the unlearned samples ~\cite{DBLP:conf/sp/BourtouleCCJTZL21,DBLP:conf/sigmod/SchelterGD21}. Therefore, the design of further unlearning schemes needs to consider that any before and after models should not expose any information about the samples that need to be unlearned. Further, the security of the data cached during the unlearning process also needs to be explored.

\subsubsection{The verification of machine unlearning}~

Verification methods should be easy to implement and applicable to users. Most current simple verification schemes, such as those based on attacks, relearning time, and accuracy~\cite{DBLP:conf/aaai/GravesNG21, DBLP:conf/cvpr/GolatkarARPS21}, are derived from existing learning or attack metrics. Those one-sided methods seldom provide strong verification of the unlearning process’s effectiveness~\cite{DBLP:conf/miccai/LiuT20, DBLP:journals/corr/abs-2003-04247, DBLP:journals/corr/abs-2110-11891}. Meanwhile, unlearning methods with a theoretical guarantee are usually based on rich assumptions and can rarely be applied to complex models since complex deep models usually make those assumptions invalid~\cite{DBLP:journals/corr/abs-2108-11577, DBLP:conf/icml/GuoGHM20}. In addition, these verification schemes are not user-friendly and easy to implement.

Therefore, the verification schemes should consider the feasibility and acceptability, that is, users should be able to understand and verify whether their unlearning request has been completed based on some simple operations. There are already some relevant schemes, such as the backdoor-based verification mechanism in~\cite{DBLP:journals/corr/abs-2003-04247} and the encryption-based verification scheme in ~\cite{DBLP:journals/tdsc/LiuMYLMR22}. However, these schemes are still quite difficult for ordinary users. Therefore, an easy-to-implement and understanding verification scheme is a topic worthy of research.

\subsubsection{The applications of machine unlearning}~

While promoting individual data privacy, machine unlearning has also gradually emerged as a solution for other applications. Regulations and privacy issues have resulted in the need to allow a trained model to unlearn some of its training data. Apart from these, there are several other scenarios where efficient machine unlearning would be beneficial. For instance, it could be used to accelerate the process of leave-one-out-cross-validation, removing adversarial or poisoning samples and identifying significant and valuable data samples within a model~\cite{DBLP:conf/nips/GinartGVZ19}. As of now, some relevant applications have emerged~\cite{DBLP:conf/infocom/LiuFCLMWM22,DBLP:journals/corr/abs-2108-11577}. For example, Alexander et al.~\cite{DBLP:journals/corr/abs-2108-11577} proposed a feature unlearning scheme that could be used to address fairness issues.

At the same time, the machine unlearning scheme can also serve as an effective attack strategy to strengthen the robustness of the model. One potential attack scenario to consider is as follows: the attacker first introduces pre-designed malicious samples into the dataset, which are subsequently used by the model provider to train the model. After that, the attacker initiates unlearning requests to remove the information about those pre-designed samples from the model, which will affect the performance and fairness of the model, or unlearning efficiency~\cite{DBLP:conf/aaai/MarchantRA22}. Therefore, in addition to strengthening data protection, machine unlearning also has enormous potential in other areas.

\subsection{Future Directions}

\textbf{Information synchronization:} Similar to process synchronization in operating systems, machine unlearning may create information synchronization problems~\cite{DBLP:journals/tods/Schlageter78,DBLP:journals/tse/Bagrodia89}. Since machine unlearning is usually computationally costly, the model provider may not be able to complete the unlearning process immediately. In the interim, how to handle incoming prediction requests deserves careful consideration. Consider that, if predictions continue to be returned prior to the model’s update, unlearned data may be revealed. However, if all requests for prediction are rejected until the unlearning process is completed, model utility and service standards will surely suffer. Therefore, how to handle prediction requests within this interval needs comprehensive consideration.

\textbf{Federated unlearning:} Federated learning is a special kind of distributed learning that is characterized by various unstable users distributed in different places, each of whom has control over their devices and data~\cite{DBLP:journals/jsac/WangTSLMHC19,DBLP:journals/comsur/KhanSHHH21}. Imteaj et al.~\cite{DBLP:journals/iotj/ImteajTWLA22} show that model providers are more likely to receive requests to remove specific samples from a model trained in a federated learning setting. For example, when a user quits the collaborative training process, they may ask for their contribution to be removed from the collaborative model. Therefore, how to efficiently realize machine unlearning in a federated learning setting, considering the limitations of such a setting, like unacceptable training data, unstable connections, etc., is worthy of research~\cite{DBLP:journals/tdsc/LiuMYLMR22}.

\textbf{Disturbance techniques:} Problems with privacy leaks before and after machine unlearning, are mainly caused by the differences between the two models. A feasible solution is to interfere with the training process or adjust the model parameters so that the model is different from what it should have been. Data disturbance techniques have the ability to interfere with specific data while ensuring overall data availability~\cite{DBLP:journals/tcss/ZhangWKJS22}. For example, Guo et al.~\cite{DBLP:conf/icml/GuoGHM20} hide information about the unlearned samples using a loss perturbation technique~\cite{DBLP:journals/jmlr/ChaudhuriMS11} at the time of training. The technique involves perturbing the empirical risk through a random linear term. As such, a useful direction for future research may be to incorporate data disturbance into machine unlearning problems and to develop new mechanisms to support more sophisticated analyses.

\textbf{Feature-based unlearning methods:} Unlearning based on model shifting usually removes the impact of the unlearning dataset by calculating the influence on the model~\cite{DBLP:conf/cvpr/GolatkarAS20, DBLP:conf/eccv/GolatkarAS20}. However, calculating the influence of the samples directly may be too complex~\cite{DBLP:conf/icml/KohL17}. Can we shift the calculation of influence from the original training samples to a group of specific features? When an unlearning request arrives, influence can be calculated based on the features instead of the original training samples. Technologies that may be relevant to this question include feature extraction~\cite{DBLP:journals/tkde/DingFZCJ22}, feature generation\cite{DBLP:journals/tip/ZhangYYWWFC22}, and feature selection\cite{DBLP:journals/tkde/ZhuLWW22}, which could be integrated into unlearning operations.

\textbf{Game-theory-based balance:} Game theory has been a booming field with several representative privacy-preserving techniques coming out in the past decade~\cite{DBLP:journals/ai/LeonardosP22}. There are many schemes involving privacy-preserving solutions based on game theory that trade-off data privacy and utility issues~\cite{DBLP:journals/jcst/CuiQNY0X19,DBLP:journals/tpds/LiangYDZ21}. For a model provider, machine unlearning is also a trade-off between model performance, and user privacy, where an over-unlearning strategy may lead to performance degradation, while insufficient protection may lead to privacy leaks. Can we formalize the unlearning problem as a game between two players: a model provider and a data provider? If so, we could provide a game model between these two entities and determine a set of strategies and utilities to figure out how to perform unlearning operations that maintain the model’s performance to the maximum extent possible. Such an approach could also protect the user’s sensitive data from being leaked. These are open issues that need to be explored further.



\section{Conclusion}
\label{sec:conclusion}

Machine learning methods have become a strong driving force in revolutionizing a wide range of applications. However, they are also bringing requests to delete training samples from models due to privacy, usability, or other entitlement requirements. Machine unlearning is a new technology that can cater to these requests for deletion, and many research studies have been carried out in this regard. In this survey, we provided a comprehensive overview of machine unlearning techniques with a particular focus on the two main types of unlearning processes: data reorganization and model manipulation. First, we provided the basic concept and different targets of machine unlearning. By analyzing typical approaches, we proposed a novel taxonomy and summarized their basic principles. We also reviewed many existing studies and discussed the strengths and limitations of those studies within each category. In addition, we emphasized the importance of verifying machine unlearning processes and reviewed the different ways in which machine unlearning can be verified. Finally, we discussed several issues that would merit future research and provided some feasible directions that need to be explored in the future. Our future work will focus on exploring the potential of machine unlearning in intriguing areas such as federated learning with a verifiability property.

     
\begin{acks}
This paper is supported in part by the Australian Research Council Discovery DP200100946 and DP230100246, and NSF under grants III-1763325, III-1909323,  III-2106758, and SaTC-1930941. 
\end{acks}

\bibliographystyle{unsrt}
\bibliography{manuscript-acmsmall}

\appendix


\end{document}